\pdfoutput=1

\documentclass[12pt,english,a4paper]{article}

\usepackage{amsmath,amsfonts,amssymb,babel,slashed,empheq,tensor}
\usepackage[utf8]{inputenc}
\usepackage{cite}
\usepackage[pdftex,breaklinks]{hyperref}
\usepackage{amsmath}
\usepackage{amsfonts}
\usepackage{array}
\usepackage{amssymb}
\usepackage{latexsym}
\usepackage{mathrsfs}
\usepackage{braket}		
\usepackage{setspace}
\usepackage{mathrsfs}
\usepackage{graphicx}
\usepackage{xcolor}
\usepackage{mathtools}
\usepackage{slashed}
\usepackage{overpic}
\usepackage[all]{xy}
\usepackage{tikz-cd}
\usepackage{slashed}
\usepackage{empheq}
\usepackage{float}
\usepackage{mathtools}
\usepackage{tikz}
\usepackage{bm}
\usepackage{pgfplots}
\pgfplotsset{compat=1.18}
\def\rlwd{.9pt}
\def\lhexbrace{\kern1pt%
\setstackgap{S}{0pt}\def\stackalignment{l}
\ThisStyle{\scalerel*{%
  \stackunder[-\rlwd]{%
    \stackon[-\rlwd]{\roundrule{\rlwd}{4pt}}{\rotatebox{60}{\roundrule{4pt}{\rlwd}}}%
  }{\rotatebox{-60}{\roundrule{4pt}{\rlwd}}}%
}{\SavedStyle[}}}
\def\rhexbrace{%
\setstackgap{S}{0pt}\def\stackalignment{r}
\ThisStyle{\scalerel*{%
  \stackunder[-\rlwd]{%
    \stackon[-\rlwd]{\roundrule{\rlwd}{4pt}}{\rotatebox{-60}{\roundrule{4pt}{\rlwd}}}%
  }{\rotatebox{60}{\roundrule{4pt}{\rlwd}}}%
}{\SavedStyle[}}\kern1pt}

\makeatletter

\newcommand{\mso}{\mathfrak{so}}

\newcommand{\hs}{\mathfrak{hs}}
\newcommand{\msu}{\mathfrak{su}}

\newcommand{\cT}{\mathcal{T}}
\newcommand{\Tr}{\text{Tr}}

\newcommand{\harold}[1]
{#1}

\newcommand{\nn}{\nonumber}
\def\obar{\overline}

 \def\one{\mbox{1 \kern-.59em {\rm l}}}

\newcommand{\cK}{\mathcal{K}}
\newcommand{\cS}{\mathcal{S}}
 \def\cQ{{\cal Q}}
 \def\cM{{\cal M}}
 \def\cN{{\cal N}}
 \def\cH{{\cal H}}
 \def\cL{{\cal L}}
 \def\cF{{\cal F}}
 \def\cC{{\cal C}
 \def\cO{{\cal O}}}
 \def\cG{{\cal G}}
 \def\cA{{\cal A}}
 \def\cO{{\cal O}}
 
 \def\cR{{\cal R}}

\def\TT{{\bf T}}

\newcommand{\Mat}{\mathrm{Mat}}
\newcommand{\tr}{\mathrm{tr}}
\newcommand{\NC}{{\rm NC}}
\newcommand{\YM}{{\rm YM}}

\newcommand{\del}{\partial}

\def\R{{\mathbb R}} \def\C{{\mathbb C}} \def\N{{\mathbb N}}

\def\a{\alpha}  \def\b{\beta}

\def\d{\delta}

\def\l{\lambda} 
\def\t{\tau} 
\def\L{\Lambda}

\newcommand{\und}{\underline}

 \allowdisplaybreaks[3]

\makeatother

\begin{document}

\renewcommand{\title}[1]{\vspace{10mm}\noindent{\Large{\bf#1}}\vspace{8mm}}
\newcommand{\authors}[1]{\noindent{\large #1}\vspace{5mm}}
\newcommand{\address}[1]{{\itshape #1\vspace{2mm}}}

\begin{titlepage}

\begin{center}

\vspace{15mm}

\title{ {\Large Quantum spacetime and gravity \\[1ex]
from the IKKT matrix model: an invitation} }

\vskip 3mm

\authors{Harold C.\ Steinacker}

\makeatletter{\renewcommand*{\@makefnmark}{}
\footnotetext{E-mail: \texttt{harold.steinacker@univie.ac.at}}\makeatother}

\vskip 3mm

 \address{
{\it Faculty of Physics, University of Vienna\\
Boltzmanngasse 5, 1090 Vienna, Austria  } }

\bigskip

\vskip 1.4cm

\textbf{Abstract}
\vskip 3mm

\begin{minipage}{14.8cm}%
\vskip 3mm

This is a concise and compact  introduction to 
a framework for spacetime, gravity, and fundamental physics based on the IKKT matrix model, focusing on conceptual and structural aspects. 
The approach rests on nontrivial vacua or matrix backgrounds, which serve as 3+1-dimensional noncommutative spacetime. We discuss
a specific class of backgrounds describing expanding homogeneous and isotropic FLRW spacetime, which provide a solution of the one-loop effective action with finitely many dof per volume.
Perturbations of the background give rise to fields propagating on spacetime governed by a weakly-coupled gauge theory, as well as a nonstandard description of (quantum) gravity leading to some IR modifications. Maximal supersymmetry of the model is essential to ensure
effective locality.

\end{minipage}

\end{center}

\end{titlepage}

\tableofcontents



\section{Introduction}

Nature is governed by quantum mechanics. This is the most profound insight of 20${}^{\rm th}$ century physics, replacing a classical picture that seemed satisfactory but is ultimately inconsistent.
Today, we still face an analogous problem with gravity:
general relativity (GR), its currently accepted description, is a classical theory that
is ultimately inconsistent and bound to fail, due to inevitable singularities of spacetime.
In a consistent theory, matter falling into a black hole should be described and accommodated in some meaningful way.
This is analogous to the classical instability of atoms, which signals the breakdown of classical physics.
Resolving these issues will likely require going beyond the classical framework of geometry.

The problem of spacetime singularities  -- as well as other issues such as black hole entropy --
 suggests that spacetime admits only finitely many degrees of freedom per volume. 
 However, the non-renormalizability of general relativity implies that different discretized versions of GR (such as loop quantum gravity \cite{Rovelli:2008zza} and spin foams \cite{Perez:2012wv}, causal dynamical triangulations \cite{Loll:2019rdj}, and many other proposals \cite{Loll:1998aj}) 
would lead to different physics, and are basically ad hoc. There might be preferred versions  as suggested by asymptotic safety \cite{Eichhorn:2018yfc,Percacci:2017fkn}, but a compelling theory ought to have a clear and simple mathematical definition.

This issue is of course addressed in string theory, which starts from an entirely different and ''special'' framework that provides a well-behaved quantum theory including gravity in 9+1 dimensions. However, standard methods to reduce it down to 3+1 dimensions introduce vast amounts of ambiguities \cite{Grana:2005jc}.
This issue is known as the landscape \cite{Susskind:2003kw}, and seems hard to accept for a fundamental theory of physics.

The purpose of this paper is to discuss the foundations of a  different approach to quantum gravity, based on a distinct matrix model of Yang-Mills-type  known as IKKT or IIB model \cite{Ishibashi:1996xs}, with action of the form $S = \tr([\TT,\TT]^2 + ...)$. While this model belongs to the broad arena of string theory and inherits much of its magic \cite{Aoki:1998bq,Fukuma:1997en}, it provides a non-perturbative and independent starting point. Its origin can be traced back to the ideas of large-N reduction \cite{Eguchi:1982nm,Gonzalez-Arroyo:1982hyq}, and it naturally arises from several different points of view\footnote{The model is nothing but the dimensional reduction of 9+1 dimensional $\cN=1$ SUSY Yang-Mills to a point, and it arises also naturally from the perspective of noncommutative gauge theory.}.

There are  different approaches even within this matrix model, leading to considerable confusion. In this paper we focus on a 
weakly coupled regime on nontrivial backgrounds, to be contrasted with holographic approaches.
In the present approach, the IKKT model leads to a nonstandard mechanism for gravity in $3+1$ dimensions, which avoids the landscape of string compactification. 
Much progress has been made in recent years, and the main hypotheses underlying this approach can now be justified to some extent. 
The aim of this paper is to resolve some of the lingering confusions, and to clarify the scope as a potential basis for fundamental physics.

From a holographic point of view following the example of AdS-CFT \cite{Maldacena:1997re}, the matrix model is viewed as a non-perturbative definition of an emergent $9+1$ dimensional target space geometry\footnote{For the BFSS model \cite{Banks:1996vh}, this leads to a relation with some black hole geometry in supergravity.}. This is relevant in the strongly coupled (deep quantum) regime and has been substantiated for the polarized IKKT model \cite{Ciceri:2025wpb,Hartnoll:2024csr,Komatsu:2024bop,Bonelli:2002mb}, associating some 9+1-dimensional supergravity backgrounds to various solutions of the matrix model. The pertinent matrix model observables in that approach are of the type $\tr(\TT....\TT)$,  viewed as geometric target space observables.

In contrast, the present approach is focused on the $3+1$-dimensional effective geometry and physics of some {\bf nontrivial matrix background $\TT$ at weak coupling}\footnote{The distinction between strongly coupled and weakly coupled regimes does make sense on a nontrivial background, even though  there is no free parameter in the IKKT model \cite{Steinacker:2026jzp}.}.
The $9+1$ dimensional target space is  unphysical in the weakly coupled regime (similar as in $\cN=4$ SYM), since all perturbation modes propagate on the 3+1-dimensional background, and cannot escape into target space. Nevertheless, 
the background might be interpreted as a brane in target space in some decoupling limit of string theory \cite{Seiberg:1999vs}. The aim is to understand the physics of fluctuations on interesting such backgrounds.

More specifically,
we consider {nontrivial 3+1-dimensional matrix backgrounds or vacua $\TT$,
where the matrix fluctuations $\TT +\cA$ are weakly coupled and  small.
 The reduction to 3+1 dimensions is hence achieved by vacuum selection, i.e. by spontaneous symmetry breaking (SSB). 
The landscape problem in string theory is thereby avoided, because there is no need for compactification; in fact compactification would destroy the non-perturbative power of the model.
This mechanism is well understood in toy models and is supported by non-perturbative studies of the IKKT model\footnote{Numerical studies are highly challenging due to the oscillatory nature of the matrix integral.} \cite{Anagnostopoulos:2026qvz,Anagnostopoulos:2026utg,Anagnostopoulos:2022dak,Nishimura:2019qal,Hatakeyama:2019jyw}, providing growing evidence for SSB towards 3+1 dimensions. Novel methods to 
tackle Yang-Mills matrix models   using quantum computers are also being developed \cite{Buckley-Bonanno:2026ygr,Rinaldi:2021jbg}. 

Instead of trying to address the difficult problem of vacuum selection within the IKKT model, 
we focus on a particularly interesting class of candidate vacua. They describe 
quantum spacetime through noncommutative algebraic structures interpreted in terms of quantized symplectic spaces.
This is familiar from quantum mechanics, where phase space is described through non-commuting matrices $X$ and $P$. In the matrix model, such backgrounds lead to a noncommutative gauge theory, which has been discussed extensively in the literature from various points of view \cite{Doplicher:1994tu,Grosse:1995ar,Szabo:2001kg,Douglas:2001ba}. Although such gauge theories generically suffer from pathological UV/IR mixing, that phenomenon is benign in the maximally supersymmetric IKKT model, and reflects the stringy nature of the theory.

In particular, 
we discuss a matrix background that describes an expanding homogeneous and isotropic 3+1 dimensional quantum spacetime with a Big Bounce, which is a solution of the one-loop effective action with remarkably nice physical properties,
including a stable dilaton and UV/IR hierarchy for a large epoch of evolution. We describe how gravity emerges as a quantum effect on this background, through a novel mechanism leading to interesting new physics that remains to be explored.

Although the present paper contains some new results and insights (notably on the massless sector  in section \ref{sec:induced-masses}  and on time \& unitarity in section \ref{sec:time-evol}), the main aim is to provide a coherent, accessible and up-to-date discussion of the framework and its basic mechanisms. 
The approach is now sufficiently developed and established to warrant an introductory overview paper.
Not all details are included for the sake of brevity, they can be found in the references, and more comprehensively in  \cite{Steinacker:2024unq}. The emphasis is on the physical mechanisms, aiming to understand if and how realistic physics may arise  from the model.

\section{Euclidean matrix models and nontrivial vacua}

This section serves as a warm-up, recalling how noncommutative (NC) gauge theory arises from Yang-Mills-type matrix models in nontrivial matrix vacua.

\subsection{The ARS model}

We start with a prototypical  matrix model in Euclidean signature, which illustrates nontrivial vacua or matrix backgrounds.
The Alekseev-Recknagel-Schomerus  (ARS) model \cite{Alekseev:2000fd} is defined by the action
\begin{align}
    S_{\rm ARS}[\TT] = N\tr \big(-\frac 14 [\TT^a,\TT^b][\TT_a,\TT_b]
    - \frac{i\alpha}{3} \varepsilon^{abc}\TT_a \TT_b \TT_c + \frac{\mu^2}{2} \TT^a \TT_a   \big)
    \label{ARS-action}
\end{align}
for 3 hermitian $N\times N$ matrices $\TT_a, \ a = 1,2,3$.
This action enjoys a "global" $SO(3)$ symmetry 
\begin{align}
    \TT^a \to  \L^a_b \TT^b 
\end{align}
 as well as a $SU(N)$ gauge invariance
\begin{align}
    \TT^a \to U^{-1} \TT^a U\ .
\end{align}
Gauge invariance means that gauge equivalent matrices  are identified. Hence bosonic physical configurations are gauge orbits of matrices $\{\TT^a\}_{/_\sim}$ denoted as {\bf matrix configurations}.

Saddle points of the action obey the classical equations of motion (eom)
\begin{align}
 \Box \TT_a =  i\a \varepsilon_{abc} \TT^b\TT^c - \mu^2 \TT_a, \qquad 
 \Box = [\TT^a,[\TT_a,\cdot]]
\end{align}
where $\Box$ is the matrix Laplacian. 
For suitable parameters, these admit fuzzy spheres with suitable radius $r$ as solution:
\begin{equation}
\label{fuzzy-S2-config}
\TT_a \;=\; r\,L_a,\qquad [L_a,L_b]=i\,\varepsilon_{abc}L_c,
\end{equation}
with $L_a$ a representation of $su(2)$. 
Stability of such a solution is obvious for a
 choice of parameters where the action takes the form
\begin{align}
S_{\rm ARS} = N\tr \Big(\big(T^a + \frac ir\varepsilon^{abc} T_b T_c\big)
\big(T_{a} + \frac ir\varepsilon_{ade} T^d T^e\big)\Big) \ 
\label{ARS-model-2}
\end{align}
and $S$ is positive definite
(this extends more generally to some domain in parameter space).
Then the global minimum $S=0$ is achieved
for any background of the form 
\begin{align}
    \TT_a = \oplus_i\, r L_a^{(N_i)}
    \label{matrix-config-S2}
\end{align}
 interpreted as a stack of fuzzy spheres $S^2_{N_i}$
with  $N_i$ with $\sum_i N_i = N$.
It is important to observe that 
the global $SO(3)$ symmetry is unbroken for these nontrivial fuzzy sphere backgrounds, because it is equivalent to a gauge transformation,
\begin{align}
\label{covariance-S2}
    \L^a_b \TT^b = U^{-1} \TT^a U
\end{align}
for $\L \in SO(3)$ with representation $U(\L) \in SU(N)$.
Such matrix configurations are denoted as {\bf covariant quantum spaces}. 

Low-energy fluctuations $\TT^a \to \TT^a + \cA^a$ on a stack of $k$ coincident fuzzy spheres  $S_{n}^2$ with $N = k n$  organize into tangential Yang-Mills $U(k)$ gauge fields on $S_{n}^2$, as discussed in more detail for $\R^2_\theta$ in the next section. Transversal fluctuations correspond to scalar fields, which plays the role of a  Higgs field if the $N_i$ are different. In addition 
 there are non-local high energy ''stringy'' fluctuation modes, which have no classical analog \cite{Steinacker:2022kji}.

\paragraph{Quantization.}

The quantization of Euclidean matrix models is  defined by integrating over the space of all matrices, which respects all symmetries. Then 
correlators are  defined as
\begin{align}
 \langle \cO(\TT)\rangle := \frac 1{Z }\int d\TT \cO(\TT) \  e^{- S[\TT]} \ , \qquad 
 Z = \int d\TT \  e^{- S[\TT]} 
 \label{correlators-Euclid}
\end{align}
where $\cO(\TT)$ is some matrix observable, i.e. some function of the matrices $\TT^a$. 
This integral is well-defined and absolutely convergent for $\mu^2 > 0$.

From the above discussion, it is evident that for suitable choices of the parameters,
there will be nontrivial matrix configurations \eqref{matrix-config-S2} of the model
which are sufficiently stable at the quantum level to support interesting physics.
This will be indicated by
\begin{align}
\boxed{ \
 \langle \TT^{a} \rangle = \obar T^{a} \
 \ }
 \label{VEV-MM}
\end{align}
where $\obar T^{a} \in \Mat(\cH)$ is some nontrivial matrix configuration.
Equivalently, the  expectation value  of fluctuations 
$\TT^a = \obar \TT^a + \cA^a$ vanishes $\langle  \cA^a \rangle = 0$, at least in a perturbative sense.
Such a matrix configuration will be considered as {\bf vacuum} or {\bf background}, similar as in quantum field theory. 
This will play a central role throughout this paper.
We will focus on the generic case of non-commutative backgrounds $[\obar T^a,\obar T^b] \neq 0$, and often drop
the bars.

In numerical studies, one typically finds a nontrivial ''fuzzy sphere'' phase with a vacuum as above, 
and a ''matrix'' phase of random matrix fluctuations around the trivial vacuum, depending on the parameters \cite{Iso:2001mg,Delgadillo-Blando:2007mqd,Delgadillo-Blando:2008cuz,Azuma:2005bj}. In the fuzzy-sphere phase, the fluctuation spectrum matches the expected decomposition into fuzzy spherical harmonics. Instabilities may arise for low harmonics, triggering collapse to the commuting phase.
An interesting question is whether there is one dominant vacuum at the quantum level, or many possible vacua at the same footing. 
Monte Carlo simulations indicate coalescence into larger irreps at sufficiently large $\alpha$, consistent with analytic one-loop calculations \cite{Azuma:2005bj}.
However, various configurations of fuzzy spheres can be  sufficiently stable to admit interesting physics \cite{Hrmo:2026ums}.

\subsection{Euclidean gauge theory from Yang-Mills matrix models}
\label{sec:Euclid-gauge-theory}

To understand the physics of the fluctuations, we consider the simpler Yang-Mills (YM)  matrix model defined by the action 
\begin{equation}
\label{YM-action}
     S_{\YM}[\TT,\Psi] =\tr \big(-\frac 1{4g^2} [\TT^a,\TT^b][\TT_a,\TT_b]
      +\overline{\Psi}\Gamma^a[\TT_a,\Psi]\big)
\end{equation}
where $\TT_a \in \Mat(\cH), \ a = 1,...,D$ are hermitian matrices transforming as vectors under $SO(D)$, 
and $\Psi$ is a matrix-valued $SO(D)$ spinor. 
Now we allow infinite-dimensional matrices, with $\cH$ a separable Hilbert space.
This action is invariant under a "global" $SO(D)$ symmetry (or rather its universal cover) acting on the indices, as well as the fundamental $U(\cH)$ gauge invariance 
\begin{align}
    \TT^a \to U^{-1} \TT^a U, \qquad \Psi \to U^{-1}\Psi U \ .
\end{align}
The model also enjoys the translation symmetry 
\begin{align}
    \TT^a \to \TT^a + c^a \one \ 
    \label{translations}
\end{align}
which may be fixed by requiring the matrices to be traceless.
The quantization is defined  by 
\begin{align}
 \langle \cO(\TT)\rangle := \frac 1{Z }\int d\TT d\Psi \cO(\TT) \  e^{- S[\TT]} \ , \qquad 
 Z = \int d\TT d\Psi \  e^{- S[\TT]} \ 
 \label{correlators-Euclid-psi}
\end{align}
where $d\Psi$ indicates a Grassmann integral.
The matrix integral 
is still well-defined under some restrictions \cite{Krauth:1998yu,Austing:2001bd}.

As in the ARS model, there are again nontrivial matrix configurations $\langle \TT^{a} \rangle = \obar T^{a} $ \eqref{VEV-MM}  of the model
at least for $N\to\infty$, 
which are sufficiently stable at the quantum level to support interesting physics. 
Such vacua should be saddle points
$\{\TT^a\}$ of the quantum effective action. 
Candidates are given by solutions of the classical matrix equations of motion 
\begin{align}
    \Box \TT^a = 0, \qquad \Box = [\TT^a,[\TT_a,\cdot]] \ 
\end{align}
where $\Box$ is the matrix Laplacian.
For finite-dimensional matrices, it is easy to see that the 
only solutions are commuting matrix configurations\footnote{This follows from the fact that $\Box$ is positive semi-definite.}, 
However, nontrivial vacua do exist\footnote{For another interesting solution see \cite{Liao:2025yfb}.} for $N=\infty$, known as Moyal-Weyl quantum plane $\R^{2n}_\theta$ 
\begin{align}
    \TT^\a = X^\a, \qquad 
    [X^\a,X^\b] = i\theta^{\a\b}\ \one \ 
   \label{MW-generic-CCR}
\end{align}
for $\a=1,...,2n$; the remaining matrices  $\TT^i$ for $i=2n+1,...,D$ vanish.
The classical eom $\Box \TT^a = 0$ is clearly satisfied. The 2-dimensional planes $\R^2_\theta$ can be considered as de-compactified fuzzy spheres, and conversely the cubic and quadratic terms in the ARS models lead to a compactification of $\R^2_\theta$ into fuzzy spheres.
If the fermionic content matches the bosonic one, the model enjoys a supersymmetry. The $\R^{2n}_\theta$ vacua then satisfy a BPS condition, and are hence protected from quantum corrections.

Hence we have again a number of different candidates for matrix vacua or backgrounds. These backgrounds can now have different dimensions $2n$, and they can be freely translated in ''target space'' $\R^D$ due to the translation symmetry \eqref{translations} of the action. 

The most ambitious goal is to understand whether there is a dominant such vacuum; this can be studied in non-perturbative approaches.
We will take a more modest {\bf perturbative} point of view, where the matrix integral is viewed -- and evaluated -- as an integral of fluctuations around the  background (or rather its gauge orbit). The background under consideration is required to be sufficiently stable in this perturbative sense.
In particular, the typical quantum fluctuations of the matrices in \eqref{correlators-Euclid-psi} should be much smaller than the
background:
\begin{align}
 \Delta^{(Q)} \cA^{a} := \sqrt{\Big\langle |\cA^{a}|^2\Big\rangle} \  & \ll |\TT^{a}|  \qquad \quad \mbox{\bf semi-classical regime} \ 
\label{class-background-crit-1}
\end{align}
for components $a$ with nontrivial background\footnote{Some refinements are required to make this criterion well-defined, see \cite{Steinacker:2026jzp}.}. In that case, noncommutative backgrounds with $[T^{a},T^{b}]\neq 0$ define a noncommutative i.e. quantum geometry.
The $U(\cH)$ gauge symmetry is then realized non-linearly on the fluctuations $\cA$, which are governed by a gauge field theory on the quantum space as discussed in section \ref{sec:NC-YM}. 
The opposite regime would be the {\bf deep quantum regime}, where quantum fluctuations dominate.
These two distinct notions of ''quantum'' on noncommutative backgrounds should not be confused.

It is not hard to see \cite{Steinacker:2026jzp} that on noncommutative backgrounds, the above criterion \eqref{class-background-crit-1} typically holds in the weak coupling regime (as defined by the background), in contrast to the strong-coupling deep quantum regime typically interpreted in terms of holography.

\section{Matrix geometry and semi-classical description}

We return to the general discussion of matrix configurations, focusing 
on non-commutative backgrounds $[\TT^a,\TT^b] \neq 0$ denoted as {\bf quantum space}. 
Given such a quantum space, the matrices are naturally organized in harmonics
$\Phi = \sum c_\L \Upsilon_\L$, defined as solutions of 
\begin{align}
\label{harmonics-scalar}
    \Box \Upsilon_\L = \lambda_\L \Upsilon_\L, \qquad \Box = [\TT^a,[\TT_a,\cdot]] \ .
\end{align}
These harmonics $\{\Upsilon_\L\}$ form an ON-basis of $\Mat(\cH)$. The tangential fluctuations $\cA^a$ can be similarly decomposed into vector-valued harmonics, in the spirit of background field theory.

For {\bf almost-commuting matrix configurations} i.e. $[\TT^a,\TT^b] \ll \TT^a \TT^b$ in some matrix norm\footnote{This can be the Hilbert-Schmidt norm, or a more local measure in terms of expectation values of quasi-coherent states; see \cite{Steinacker:2020nva,Steinacker:2026jzp} for a more detailed discussion.},
the algebra of matrices can typically be related to the algebra of functions on some underlying manifold $\cM$,
\begin{align}
    \Mat(\cH) \ &\sim  \  \cC(\cM)   \nn\\
    \Phi \ &\sim \ \phi \nn\\
       [\cdot,\cdot] & \sim  i\{\cdot,\cdot\} \nn\\
       \Tr(\Phi) &\sim \int \Omega \phi \ .
       \label{semi-classical}
\end{align}
Then commutators reduce to Poisson brackets, which define a symplectic structure in the non-degenerate case, with symplectic volume form $\Omega$. This {\em semi-classical} or Poisson description is typically accurate for long wavelengths, above some scale  $L_\NC$ of noncommutativity.

Moreover, any matrix background
$\TT^a$ defines a set of derivations
\begin{align}
    e^a = -i[\TT^a,\cdot] 
    \label{frame-matrix}
\end{align}
on the space of matrices $\Mat(\cH)$, which  respect the (Hilbert-Schmidt) inner product $\tr(A^\dagger B)$. Hence these vector fields 
define unitary 1-parameter evolution operators
   $e^{-it[\TT^a,\cdot]}$
acting on the space of matrices $\Mat(\cH)$. 
In the semi-classical regime,
these derivations $e^a$ define a set of canonical vector fields on $\cM$,
\begin{align}
    e^a = -i[\TT^a,\cdot] &\sim  \{\TT^a,\cdot\} 
    \label{frame-1} \\[1ex]
    e^{a\mu} &\sim  \{\TT^a,x^\mu\}   \nn
\end{align}
which play the role of a {\bf frame\footnote{An early precursor was proposed in \cite{Madore:2000aq}, however with algebraic constraints. The frame arises automatically from the matrix-model framework, without these constraints.}}; here $x^\mu$ are local coordinate functions on $\cM$.

On the other hand, due to the relation $\Mat(\cH) \sim  \cC(\cM)$ the 
 matrices $\TT^a$ can also be interpreted as 
quantized embedding functions 
\begin{align}
    \TT^a \sim {\bf t^a}: \quad \cM \hookrightarrow \R^{D}
\end{align}
of $\cM$ into target space.
These two interpretations of the matrices are perfectly consistent: observables in classical or quantum mechanics can be viewed either as functions or as generators of canonical transformations.
Both points of view provide complementary insights.

As usual in gravity, the  frame determines
(up to a conformal factor)
 an effective metric, which governs the kinetic term of all fluctuation modes on the background
\cite{Steinacker:2010rh,Sperling:2019xar,Steinacker:2024unq}:
\begin{align}
G_{\mu\nu} &:= \rho^2 \gamma_{\mu\nu}, 
\qquad \gamma^{\mu\nu} = \eta_{ab} e^{a \mu}e^{b \nu}  \label{eff-metric-def}\\
 \rho^2 &= \rho_M \sqrt{|\gamma^{\mu\nu}|}
  = \frac{\sqrt{|G|}}{\rho_M} \ . 
  \label{dilaton-def}
\end{align}
Here $\Omega = d^4 x\rho_M$ is the symplectic volume form on $\cM$, which we assumed to be 4-dimensional to be specific. 
The dilaton measures the number of degrees of freedom per Riemannian volume.
This it is easily derived from the quadratic action governing the transversal matrices $T^i \equiv \phi$, which play the role of scalar fields; $\rho^2$ is determined by requiring that 
\begin{align}
 S[\phi] =  \Tr([\TT^{a},\phi][\TT_{a},\phi]) 
  &\sim -\int\Omega\{\TT^a,\phi\}\{\TT_a,\phi\} 
   = - \int d^4x\,\rho_M \gamma^{\mu\nu}\del_\mu\phi\del_\nu\phi \nn\\
  &\stackrel{!}{=} -\int d^4x\,
  \sqrt{|G_{\mu\nu}|} G^{\mu\nu}\del_\mu\phi\del_\nu\phi \ .
  \label{S-kin-phi-general}
 \end{align}
That metric turns out to
 govern the kinetic action for {\em all} fluctuations on the matrix background, therefore it must be interpreted in terms of gravity. The metric is also encoded in the matrix Laplacian (or d'Alembertian)
 \begin{align}
     \Box = [\TT^a,[\TT_a,\cdot]] \sim \rho^2 \Box_G
     \label{Box-metric-Laplacian}
 \end{align}
 that governs the low-energy physics,
 where $\Box_G$ is the Laplace-Beltrami operator for $G_{\mu\nu}$.
 The matrix harmonics \eqref{harmonics-scalar} thus becomes eigenfunctions of $\rho^2\Box_G$ on $\cM$.
For example,
the harmonics of $\Box$ for the background \eqref{fuzzy-S2-config} are  known as fuzzy spherical harmonics.

In Euclidean signature, the effective metric can also be extracted from a Connes-type distance formula \cite{connes1994noncommutative} for $\slashed{D} = \Gamma^a [\TT_a,\cdot]$.

A matrix configuration provides further structure, notably a notion of optimally localized quasi-coherent states $|x\rangle$ \cite{Steinacker:2020nva}, defined as ground states of the displacement Hamiltonian
\begin{align}
\label{displacement-H}
    H_x = \sum_a (\TT_a - x_a\one)^2, \qquad H_x  |x\rangle = \l(x) |x\rangle
\end{align}
for $x\in \R^D$.
These states allow to extract local properties such as fields $\phi(x) = \langle x|\Phi|x\rangle$, and to reconstruct $\cM$ with its Berry connection $iA=\langle x|d|x\rangle$ and symplectic structure $\omega = dA$ from the set of quasi-coherent states for a given matrix background \cite{Steinacker:2020nva}. This provides useful 
information about the local structure of a quantum space, as well as topological invariants such as Chern numbers
\begin{align}
\label{chern-c1}
  c_1 :=  \int_{S^2} \frac{\omega}{2\pi} \ .
\end{align}
This recovers e.g. the index $n$ of a fuzzy sphere $S^2_{n+1}$ from its matrix configuration, but applies to more general settings \cite{Abanov:2025lln}. 
In the present context,
quasi-coherent states are very useful to evaluate the trace $\Tr_{\Mat(\cH)}$, and hence the one-loop effective action \eqref{Gamma-IKKT} \cite{Steinacker:2024unq,Steinacker:2022kji}.

\subsection{Noncommutative Yang-Mills in the example of $\R^4_\theta$}
\label{sec:NC-YM}

We can now understand the physics of the low-energy fluctuations in the Euclidean Yang-Mills matrix model.
The Moyal-Weyl background \eqref{MW-generic-CCR}  admits a semi-classical description \eqref{semi-classical} with a constant Poisson tensor $\theta^{\mu\nu}$, via the (Weyl) quantization map
\begin{align}
    \cQ:\quad \cC(\R^4) & \to \Mat(\cH) \nn\\
      e^{i k x} &\mapsto e^{i k X} \ .
\end{align}
Functions in $X^\mu$ are thereby interpreted as scalar fields on $\R^{4}_\theta$. 
The background
\begin{align}
\obar T^{a} = 
\begin{pmatrix}
    X^\a \\ 0
\end{pmatrix} , \quad \a=1,...,4 \ 
\end{align}
defines a frame 
\begin{align}
    e^{\a\mu} = \theta^{\a\mu} \ .
\end{align}
To understand the fluctuations $\TT^{a} = \obar T^{a} + \cA^{a}$
on this background, we rewrite them  as 
\begin{align}
\cA^{\dot a} = 
\begin{pmatrix}
    \cA^\a \\ \cA^i
\end{pmatrix} = \begin{pmatrix}
   e^{\a\mu} A_\mu \  \\  \phi^i
\end{pmatrix} .
\label{A-tang-trans}
\end{align}
One then finds using $[X^\a,\cdot] = ie^{\a\mu}\partial_\mu$
\begin{align}
[\TT^\a,\phi] &= [X^\a + \cA^\a,\phi] = i e^{\a\mu}
(\partial_\mu \phi - i [A_\mu,\phi]) \, =: \,  i e^{\a\mu} D_\mu \phi, \nn\\
\,[\TT^\a,\TT^\b] &=: - i\cF^{\a\b}  = ie^{\a\mu} e^{\b\nu}
(-\theta^{-1}_{\mu\nu} + F_{\mu\nu})
\label{XX-gauge}
\end{align}
where $F_{\mu\nu} = \partial_\mu A_\nu - \partial_\nu A_\mu - i [A_\mu,A_\nu]\,$ is the
$U(1)$ field strength on $\R^4_\theta$.
Then the bosonic action takes the semi-classical form of a 
noncommutative Yang-Mills gauge theory on $\R^4_\theta$
\begin{align}
S[A,\phi] 
&= \int\limits_{\R^{4}_\theta} \frac{d^{4} x}{(2\pi)^2}\,\sqrt{|G|}
\Big(- \frac{1}{g_\YM^2} \,F_{\mu\nu}\,F^{\mu\nu}
\, - 2 D^\mu\phi^i D_\mu \phi_i
 + {g_\YM^2} [\phi^i,\phi^j][\phi_i,\phi_j] \Big) 
\label{action-YM-scalars-R2n}
\end{align}
where covariant indices are contracted with $G_{\mu\nu}$, and the effective coupling
constant 
\begin{align}
    g^2_\YM = g^2\rho^{-2} \ 
    \label{YM-coupling-MW}
\end{align}
is assumed to be small.
The gauge symmetry 
acts non-linearly on the fluctuations as 
\begin{align}
A_\mu \to U^{-1} A_\mu U + i U^{-1} \partial_\mu U, \qquad
\phi^i \to U^{-1} \phi^i U \ .
\label{gauge-A-phi}
\end{align}
 This extends straightforwardly to a 
$U(k)$ Yang-Mills theory upon replacing the background by a stack of coinciding $\R^4_\theta$ ''branes''
\begin{align}
\label{nonabel-background-MW}
\obar T^{a} = 
\begin{pmatrix}
    X^\a\otimes \one_k \\ 0
\end{pmatrix} \ .
\end{align}
Despite appearance, the $U(1)$ sector of this theory -- which would be free on ordinary spacetime -- is geometrical, since it changes the background and the effective metric $G_{\mu\nu}$. 
Space(time) and fields are thus treated on the same footing in matrix models.

\section{The Lorentzian IKKT model}

In order to obtain spacetime, we must embrace Lorentzian matrix models, and notably the Lorentzian IKKT model which is defined by \cite{Ishibashi:1996xs}
\begin{align}
   \label{IIB-action}
S_{\rm IKKT} = \mbox{Tr}\left(\frac 14[\TT_{a}, \TT_{b}] [\TT^{a}, \TT^{b}]
+  \bar{\Psi}\Gamma^{a}[\TT_{a}, \Psi]\right) \ 
\end{align}
where $\TT_{a},\  a = 0,...,9$ are Hermitian matrices, and $\Psi$ are matrix-valued Majorana-Weyl spinors of  $SO(9,1)$. \harold{We omit a coupling constant in front of the action, since it can always be absorbed by rescaling the matrices. Hence the matrices are considered as dimensionless, and dimensionful quantities can only arise on some given matrix background.}

The quantization -- defined by integrating over the space of all matrices -- now amounts to an oscillatory integral, which requires regularization through a suitable $i\varepsilon$ prescription indicated by $S_{\varepsilon}$. Then 
correlators are then defined as
\begin{align}
 \langle \cO(T)\rangle := \frac 1{Z_\varepsilon }\int dT d\Psi \cO(T) \  e^{i S_\varepsilon[T]} \ , \qquad 
 Z_\varepsilon = \int dT d\Psi \  e^{i S_\varepsilon[T]} \ .
 \label{correlators-Mink}
\end{align}
One way to approach this is via a 
Wick rotation such as $\TT^0 \to e^{i3\pi/8} \TT^{0},\ \TT^i  \to  e^{-i\pi/8} \TT^{i}$ (along with a suitable phase factor for the action, cf. \cite{Anagnostopoulos:2026qvz}).
This is useful notably for the trivial vacuum, and it allows to
relate Euclidean observables\footnote{Such target space observables are the focus of the holographic approach, aiming to extract an effective target space metric at strong coupling.} such as $\Tr(\TT_a \TT^b)$  to  Wick-rotated Lorentzian observables via contour integration. However, a Wick rotation  does not preserve hermiticity of the matrices, and it  misses the  Lorentzian saddle points of interest\footnote{Euclidean saddle points  would correspond to instanton-type configurations in the Lorentzian models.} 
Therefore a full Wick rotation is not useful for the present approach. 

A better way to define the Lorentzian model is via an  ''infinitesimal Wick rotation''
\begin{align}
\label{infinites-Wick}
    \TT^0 \to e^{i\varepsilon} \TT^{0},\quad \TT^i  \to  e^{-i\varepsilon} \TT^{i}
\end{align}
 or similar. This leads to Feynman's $i\varepsilon$ prescription for the emergent field theory on Lorentzian saddle points, and it contributes exponential damping to the matrix integral.
However, the action still has flat directions for commuting matrices.
We remove those by adding an extra regularizing mass term 
\begin{align}
\label{S-mink-reg-mass}
    S_\varepsilon = S + \tr \Big( i \varepsilon \sum_{a} (\TT^a)^2 \Big)
\end{align}
 slightly breaking the global $SO(9,1)$. 
This leads to an explicit Gaussian damping of the Lorentzian matrix integral which is now absolutely convergent, even in the presence of the Pfaffian obtained by integrating out the fermionic matrices.
The Lorentzian model is then no longer related to the Euclidean model by Wick rotation. 
Including a (small) mass term may also help to select a particular (type of) saddle point\footnote{Another interesting approach is to "gauge-fix" the non-compact $SO(9,1)$ symmetry \cite{Asano:2024def}}, cf. \cite{Anagnostopoulos:2026qvz}.

\subsection{Vacua, frame and geometry}

We follow again a perturbative approach, 
focusing on some nontrivial matrix configuration  $\langle \TT^{a} \rangle = \obar T^{a}$  \eqref{VEV-MM} of the model which is sufficiently stable for large $N =\dim\cH$.
Such vacua are known to exist in the $N\to\infty$ limit, where
the IKKT model admits many different saddle points such as $\R_\theta^{2n}$.
It is thus plausible that for large but finite $N$, the model admits -- either at the quantum level or upon adding a small mass term --- a number of distinct saddle points or vacua. 
We want to understand the physics of fluctuations around such a vacuum. 

Again,
nontrivial vacua should be saddle points
$\{\TT^a\}$ of the quantum effective action. 
Candidates for such backgrounds are given by solutions of the classical  equations of motion 
\begin{align}
    \Box \TT^a = 0, \qquad \Box = [\TT^a,[\TT_a,\cdot]]  
     \label{Box-metric-dAlembertian}\ .
\end{align}
For finite-dimensional matrices, these equations have no solutions\footnote{This follows from the fact that $\Box$ is positive semi-definite if acting on $T^0$, cf. \cite{Steinacker:2017vqw}.}, but solutions do exist\footnote{for example, $\TT^0 = \sqrt{2} \l_4, \ \TT^1 = \l_1, \ \TT^2 = \l_2$
for $\msu(3)$ generators $\l_a$ 
gives a solution with $m^2 = 2$.}  for the mass-deformed equations of motion with positive mass,
\begin{align}
    \Box \TT^a = \mu^2 \TT^a , \qquad \mu^2 > 0 \ .
\label{Box-eigenvalues-pos}
\end{align} 
In the undeformed IKKT model with $\mu=0$, 
such Lorentzian saddle points are expected to be stabilized by quantum effects \cite{Manta-new}.

For $N = \infty$, one vacuum of the IKKT model is given by $\R^4_\theta$ \eqref{nonabel-background-MW} or rather $\R^{3,1}_\theta$,
leading to maximally supersymmetric noncommutative $\cN=4$ SYM as in section \ref{sec:NC-YM}; that background is hence BPS protected. Analogous statements hold for   $\R_\theta^{2n}$ for any $2n\leq 10$. 
This leads to an important corollary: at the non-perturbative level, noncommutative $\cN=4$ SYM on $\R^4_\theta$ is {\bf the same theory} as the maximally SUSY gauge theory on any $\R^{2n}_\theta$ which arises from the IKKT model on that background, for $N=\infty$. This fact is not visible in the perturbative weak coupling regime of the gauge theory\footnote{This richer vacuum structure also explains the point-like extreme UV solitons in $\cN=4$ NC SYM known as fluxons \cite{Douglas:2001ba}.}, analogous to the different vacua encountered in the ARS model. It is also hidden in the star-product formalism of NC gauge theory.

Another important observation is that fluctuations around a vacuum describing some quantum space $\cM$ (e.g. with $\dim\cM = 3+1$) are automatically confined to $\cM$ at weak coupling. Indeed
all matrix fluctuations play the role of (local or non-local) functions on $\cM$ due to \eqref{semi-classical}, and  cannot escape into the $9+1$-dimensional target space outside of $\cM$.
These fluctuation modes are governed by a Yang-Mills type action, 
and the matrix integral \eqref{IIB-action}  amounts to a path integral quantization of that field theory, with weight $e^{\frac{i}{\hbar}S}$.
Since the IKKT model is related to IIB string theory,
this mechanism provides an alternative to
string compactifications, and thus avoids their lack of predictivity.

On the other hand, it is not clear whether one saddle point will dominate the matrix integral, or many different saddle points (possibly with different dimensions) contribute significantly, with complex weights. 
Although this is not essential for the  physics on some background of interest, it would be desirable to show that the ''dominant'' vacua are  3+1 dimensional. There is growing evidence from numerical simulations that this is indeed the case \cite{Anagnostopoulos:2026qvz,Anagnostopoulos:2026utg,Anagnostopoulos:2022dak,Nishimura:2019qal,Hatakeyama:2019jyw}; for attempts to understand this analytically see e.g.\cite{Brandenberger:2024ddi,Brahma:2022dsd,Kawai:2002jk,Aoyama:2010ry}.
From a perturbative point of view, it is clear that an almost-local quantum effective action can arise only for backgrounds with dimension $\leq$ 4. In higher dimensions, the YM theory would be UV divergent, leading to pathological non-locality in the effective action due to UV/IR mixing via non-local string modes  $|x\rangle\langle y|$ running in the loops. 
This also suggests an instability of the background.
On the other hand,
spacetimes with dimension $<4$ are expected to be irrelevant under the matrix integral.
However, settling this question will presumably require  new techniques, cf. \cite{Koch:2021yeb,Maeta:2026oku,Maeta:2026miu} for possibly relevant developments.

\subsection{Geometric structure: Weitzenböck connection and torsion}

From now on we focus on 3+1 dimensional backgrounds $\TT^{\dot\a}$. To avoid confusion with coordinate indices, we will often denote the $3+1$ matrix or frame indices with dotted greek letters, indicating that they transform as vectors under the global $SO(3,1)$ symmetry.

As in the Euclidean case, the background defines  a frame $e^{\dot\a} = \{\TT^{\dot\a},\cdot\}$ \eqref{frame-1} in the semi-classical regime, 
and thereby an effective metric.
This frame is divergence-free \eqref{div-free} due to the Jacobi identity, 
hence there is no frame bundle. Instead of working with the Levi-Civita connection, it is then more natural to work with the unique connection which respects the given frame, 
known as Weitzenböck connection
\begin{align}
    \nabla^{(W)} e^{\dot\a} = 0 \ .
\end{align}
This connection is compatible with \harold{the metric $\gamma_{\mu\nu} = \rho^{-2} G_{\mu\nu}$ \eqref{eff-metric-def}}  and flat, but its torsion encodes the same information as the standard spin connection. The torsion can be expressed via the bracket structure as follows \cite{Steinacker:2020xph}
\begin{align}
 \cT[ e_{\dot\a},e_{\dot\b}] &= \nabla^{(W)}_{\dot\a} e_{\dot\b} - \nabla^{(W)}_{\dot\b}  e_{\dot\a} 
   - [e_{\dot\a}, e_{\dot\b}] 
  = - [e_{\dot\a},e_{\dot\b}] \nn\\
  &= -\{\TT_{\dot\a},\{\TT_{\dot\b},\cdot\} \} + \{\TT_{\dot\b},\{\TT_{\dot\a},\cdot\} \}   \nn\\[1ex]
  &= \{\cF_{{\dot\a}{\dot\b}},\cdot\} , 
  \qquad\qquad 
  \cF_{\dot\a\dot\b} = i [\TT_{\dot\a},\TT_{\dot\b}] \sim  -\{\TT_{\dot\a},\TT_{\dot\b}\} 
 \label{torsion-explicit}
\end{align}
using the Jacobi identity, hence
\begin{align}
\label{torsion}
    \cT^{\dot\a\dot\b\mu} =\{\cF^{\dot\a\dot\b},x^{\mu}\} 
    \ = e^{\dot\b\nu}\del_\nu e^{\dot\a\mu} - e^{\dot\a\nu}\del_\nu e^{\dot\b\mu} \ .
\end{align}
That formula underlies the importance of the Weitzenböck connection in matrix models, and it is crucial to obtain the induced Einstein-Hilbert action \eqref{Gamma-EH-0}.

For covariant quantum spacetime discussed below,
the dimension of $\cM$ will be larger than the number of background matrices, leading to a higher-spin structure of the torsion.

\subsection{Covariant quantum spacetime}

Covariant quantum spaces or spacetimes are matrix backgrounds which enjoy covariance in the sense of \eqref{covariance-S2}, under a sufficiently large and interesting covariance group. Several interesting candidates have been discussed in the literature, including fuzzy $S^4_N$, the fuzzy hyperboloid $H^4_n$, and covariant quantum spacetime $\cM^{3,1}_n$. We will focus on the latter because it provides a near-realistic candidate for physical spacetime, albeit with $\dim\cH = \infty$. 
\harold{Before giving the precise group-theoretical construction, we briefly summarize the main geometric objects and their physical significance:} 

\harold{We will consider a specific matrix background defined by 3+1 matrices or operators $\TT^{\dot\a}$, 
based on a specific unitary representation of $\mso(4,2)$. The space of all operators on the underlying Hilbert space -- hence the space of all modes of the resulting gauge theory -- can be interpreted as quantized space of functions on twistor space $\C P^{1,2}$. This 6-dimensional symplectic space is viewed as a bundle $\cM$ over the base manifold $\cM^{3,1}$, interpreted as spacetime. 
The fiber $S^2$ transform under rotations of spacetime, so that the harmonics on $S^2$ play the role of higher-spin $\hs$ modes on spacetime. All fluctuation fields on the background can thus be organized as $\hs$-valued eigenmodes of the matrix d'Alembertian $\Box$, which governs the kinematics of the resulting higher spin gauge theory.
Spacetime $\cM^{3,1}$ is seen to have the geometry of an asymptotically flat $k=-1$ FLRW spacetime, and
some of the $\hs$ fluctuations  on spacetime will be responsible for the gravitational sector of the theory.}

\subsubsection{Rigid cosmological background}

Now we discuss an interesting (prototype for a) matrix background denoted as covariant quantum spacetime. Let $M^{AB}, \ A,B=0,...,5$ be the hermitian generators of $\mso(4,2) \cong \msu(2,2)$
in the minimal unitary ''doubleton'' representation\footnote{This representation can be explicitly constructed in terms of 4 harmonic oscillator algebras. There is also a series of generalized doubleton irreps $\cH_n$ for $n\in\N$ with very similar properties, cf. \cite{Sperling:2018xrm,Fernando:2009fq}.} $\cH_0$, and define
\begin{align}
    T^\mu = r^{-1} M^{\mu 4}, \qquad  X^a = r M^{a 5}  
\end{align}
for $\mu=0,...,3$ and $a=0,...,4$.
\harold{The signature convention is $(-++++-)$}.
Due to the special choice of representation, these generators satisfy the constraints
\begin{subequations}
\label{constraints-CP12}
    \begin{align}
    X_\mu X^\mu  &=  r^2 \one - X_4^2 \,,   \label{radius-constraint}\\
    T_\mu T^\mu &= -\frac 1{r^2} \one + \frac{X_4^2}{r^4} \,,\label{Tsquare-id}\\
    X_\mu T^\mu + T_\mu X^\mu &= 0 \ . \label{TX-id}
\end{align}
\end{subequations}
The space of matrices $\cC := \Mat(\cH_0)$
can be identified as (functions on) quantized twistor space $\cM \cong\C P^{1,2}$, which is a coadjoint orbit of $SU(2,2)$. It
decomposes into $\hs$ subsectors 
\begin{align}
\label{Cs-sectors-NC}
    \Mat(\cH_0) =: \cC =  \bigoplus_{s=0}^\infty \cC^s \ \sim \cC(\cM) \ .
\end{align}
Here 
$\cC^s$ denotes the eigenspace of an $SO(4,1)$ Casimir $\cS^2$ with eigenvalue $2s(s+1)$ \cite{Sperling:2019xar}; in particular, $\cC^0$ can be identified with completely symmetrized functions of $X^\mu$ \cite{Sperling:2018xrm}.
This quantum space has {\bf finitely many degrees of freedom per volume}, which follows from the discrete structure of the lowest-weight Hilbert space $\cH_0$ shown in figure \ref{fig:weights-minirep}.
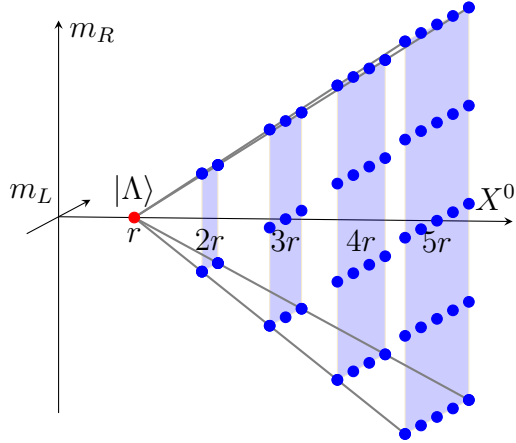
\begin{figure}[h!]
\begin{center}
\begin{tikzpicture}
    \begin{axis}[
        view={8}{5}, 
        xlabel={$X^0$},
        ylabel={$m_L$},
        zlabel={$m_R$},
        grid=both,
        axis lines=middle,
        xmin=0, xmax=6,
        ymin=-4, ymax=4,
        zmin=-4, zmax=4,
        ticks=none
    ]
    \newcommand{\drawlayer}[2]{%
        \pgfmathsetmacro{\halfsize}{#2 - 1}
        \pgfmathsetmacro{\step}{2 * \halfsize / (#2 - 1)}
        \pgfmathsetmacro{\numDivisions}{#2 - 2} 

        \foreach \i in {0,...,#2} {%
            \ifnum\i<#2 
                \pgfmathsetmacro{\pos}{-\halfsize + \i * \step}
                \addplot3[only marks, mark=*, color=blue] coordinates {(#2, \pos, -\halfsize)};
                \addplot3[only marks, mark=*, color=blue] coordinates {(#2, \pos, \halfsize)};
                \addplot3[only marks, mark=*, color=blue] coordinates {(#2, -\halfsize, \pos)};
                \addplot3[only marks, mark=*, color=blue] coordinates {(#2, \halfsize, \pos)};
            \fi
            
        }
        \foreach \i in {1,...,\numDivisions} { 
            \foreach \j in {1,...,\numDivisions} { 
                \pgfmathsetmacro{\posX}{-\halfsize + \i * \step}
                \pgfmathsetmacro{\posY}{-\halfsize + \j * \step}
                \addplot3[only marks, mark=*, color=blue] coordinates {(#2, \posX, \posY)};
            }
        }

        \addplot3[patch, patch type=rectangle, fill=blue, opacity=0.2] coordinates {
            (#2, -\halfsize, -\halfsize)
            (#2, -\halfsize,  \halfsize)
            (#2,  \halfsize,  \halfsize)
            (#2,  \halfsize, -\halfsize)
        };
    }

    \newcommand{\connectlayers}[2]{%
        \pgfmathsetmacro{\halfsizeA}{#1 - 1}
        \pgfmathsetmacro{\halfsizeB}{#2 - 1}
        \addplot3[thick, color=gray] coordinates {(#1, -\halfsizeA, -\halfsizeA) (#2, -\halfsizeB, -\halfsizeB)};
        \addplot3[thick, color=gray] coordinates {(#1, -\halfsizeA,  \halfsizeA) (#2, -\halfsizeB,  \halfsizeB)};
        \addplot3[thick, color=gray] coordinates {(#1,  \halfsizeA, -\halfsizeA) (#2,  \halfsizeB, -\halfsizeB)};
        \addplot3[thick, color=gray] coordinates {(#1,  \halfsizeA,  \halfsizeA) (#2,  \halfsizeB,  \halfsizeB)};
    }

    \addplot3[only marks, mark=*, color=red] coordinates {(1, 0, 0)}; 
    \node at (axis cs:1, 0, 0) [anchor=south] {$\ket{\Lambda}$}; 
    
    \drawlayer{2}{2} 
    \node at (axis cs:2, 0, -1) [anchor=south] {}; 
    
    \drawlayer{3}{3} 
    \drawlayer{4}{4} 
    \drawlayer{5}{5} 

    \connectlayers{2}{3}
    \connectlayers{3}{4}
    \connectlayers{4}{5}

    \pgfmathsetmacro{\halfsizeTwo}{2 - 1}
    \addplot3[thick, color=gray] coordinates {(1, 0, 0) (2, -\halfsizeTwo, -\halfsizeTwo)};
    \addplot3[thick, color=gray] coordinates {(1, 0, 0) (2, -\halfsizeTwo,  \halfsizeTwo)};
    \addplot3[thick, color=gray] coordinates {(1, 0, 0) (2,  \halfsizeTwo, -\halfsizeTwo)};
    \addplot3[thick, color=gray] coordinates {(1, 0, 0) (2,  \halfsizeTwo,  \halfsizeTwo)};
    \node at (axis cs:1, 0, 0) [anchor=north] {$r$};
    \node at (axis cs:2, 0, 0) [anchor=north] {$2r$};
    \node at (axis cs:3, 0, 0) [anchor=north] {$3r$};
    \node at (axis cs:4, 0, 0) [anchor=north] {$4r$};
    \node at (axis cs:5, 0, 0) [anchor=north] {$5r$};

    \end{axis}\label{fig:weights}
\end{tikzpicture}
\end{center}
\caption{Weight structure of $\cH_0$, with $m_{L,R}$ denoting  $SU(2)_L$ and $SU(2)_R$ weights.}
\label{fig:weights-minirep}
\end{figure}
For example, the  number of states in 
$\cH_{0,N} := \{|\psi\rangle\in\cH_0; \ X^0 \leq Nr\}$
with $X^0 \leq N r$ is given by 
the symplectic volume of the corresponding subspace of $\cM$
\cite{Manta:2025inq}
\begin{align}
\label{cH-R-dim}
 \dim \cH_{0,N}  \sim  \frac 13 N^3 
    \sim \int\limits_{x_0 \leq Nr} \Omega \ 
\end{align}
for large $N$.
This implies also an effective upper limit
for the spin $s$ of localized $\hs$ modes
\begin{align}
\label{hs-cutoff}
  s \leq s_{max} = r^{-1} X^0 \ .
\end{align}
The commutation relations\footnote{Although the $X^\mu$ and $T^\nu$ are $\mso(4,2)$ Lie algebra generators, the closed form of the rhs depends on the doubleton representations $\cH_n$ under consideration, and is simplest for $n=0$.} for the minimal $n=0$ case can be written in closed form 
\begin{subequations}
\label{minimal-cov-algebra}
\begin{align}
[T^\mu,X^\nu] &= \frac ir X_4 \,\eta^{\mu\nu} \\
 [X^\mu,X^\nu] &= -ir^3 X_4^{-1}(T^\mu X^\nu - T^\nu X^\mu)\ ,  \\
 [T^\mu,T^\nu] &= i r^{-1} X_4^{-1}(T^\mu X^\nu - T^\nu X^\mu) \ ,
  \label{XT-CR}
\end{align}
\end{subequations}
 where $X_4^2 = r^2 -  X_\mu X^\mu$.
The $T^\mu$ are eigenvectors of $\Box_T = [T^ \mu,[T_\mu,\cdot]]$ with positive eigenvalue
\begin{align}
\label{Box-T}
    \Box T^\mu = \frac{3}{r^2} T^\mu \ ,
\end{align}
which can be viewed as equations of motion of the matrix model with mass term. 
Note that a positive eigenvalue of $\Box$
indicates that the background $T^\mu$ should admit finite-dimensional deformations with similar structure, according to the discussion around \eqref{Box-eigenvalues-pos}.
This is not be the case for the  background $X^\mu$,  which satisfies $\Box_X X^\mu = - 3 r^2 X^\mu$.

We thus consider 
the following matrix background for the IKKT model
\begin{align}
    \TT^{\dot\a} = T^{\dot\a}, \qquad {\dot\a} = 0,...,3 \ ,
\end{align}
which  is very interesting for several reasons. Most importantly, $SO(3,1)$ covariance
\begin{align}
    \Lambda^{\dot\a}_{\ \dot \b} \TT^{\dot\b} = U^{-1} \TT^{\dot\a} U
    \label{covar-BG}
\end{align}
 means that $SO(3,1)$ is in fact unbroken in the matrix model. This will imply exact homogeneity and isotropy, even though there are only finitely many degrees of freedom per volume.
 That background will be slightly generalized below to accommodate the dynamics, and
 the uniqueness statement in section \ref{sec:dynamical-covar} will imply the existence of such a solution at the quantum level. 
The construction of quasi-coherent states \eqref{displacement-H} on $\cM$ is somewhat subtle, and we refer to \cite{Manta:2025inq} for details.

\paragraph{Bundle structure over spacetime.}

In the semi-classical regime,
the $X^\mu\sim x^\mu$ can be viewed as bundle \harold{projection}
\begin{align}
\label{x-projection-map}
     x^\mu:\ \cM \to \cM^{3,1} \ \subset \R^{3,1}
\end{align}
compatible with $SO(3,1)$.
\harold{The base manifold $\cM^{3,1}$ plays the role of spacetime.}
The fiber of this projection is resolved by the $t^\mu \sim T^\mu$ generators, which describe a space-like $S^2$.
This can be seen by evaluating the  constraints 
\begin{align}
   r^2 t_\mu t^\mu &= - r^{-2} x_\mu x^\mu \ , \qquad
  t_\mu x^\mu = 0 \quad  
    \label{TT-XX-relation-2}
\end{align}
near some point $\xi^\mu = (\xi^0,0,0,0)$ on the base manifold $\cM^{3,1}$. \harold{$SO(3,1)$ covariance  (i.e. equivariance) of the map 
\eqref{x-projection-map} implies that} the $S^2$ fiber rotates under the local $SO(3)$ stabilizer of any point of  $\cM^{3,1}$.
Therefore harmonics on $S^2$ play the role of higher-spin ($\hs$) modes on spacetime.
The decomposition \eqref{Cs-sectors-NC} of the algebra of functions $\cC$ into 
$\hs$ modes can now be made explicit
\begin{align}
\label{hs-decomp-spacelike}
    \cC = \C[[x^\mu,t_\nu]]/_\sim \
    = & \ \  \cC^0 \ \oplus \ \cC^1 \ \oplus \ \cC^2
 \oplus ...  \nn\\
 & \ \ \cC^s \ni \phi^{(s)} = \phi_{\und{\mu}}(x) u^{\und{\mu}}
\end{align}
in terms of irreducible polynomials in $u^\mu = \frac{r^2}{x_4}t^\mu$ of degree $s$, which generate an internal unit sphere $S^2$.
\harold{They can be decomposed further  into eigenmodes of the matrix d'Alembertian $\Box_T$ \cite{Steinacker:2024unq,Manta-new}
 \begin{equation}
    \Box_T\phi^{(s)}
    \simeq
    \big((\Box_T + m_s^2)\phi_{\und{\mu}}
    \big)u^{\und{\mu}} \, ,
    \label{hs-eigenmodes}
\end{equation}
with $\hs$ mass $m^2_s \sim  \frac{s}{16r^2}$ set by the cosmic curvature scale \eqref{a(t)-explicit}.}
 The $u^\mu$ are space-like $u_\mu u^\mu = 1$
due to the constraint $t_\mu x^\mu = 0$, so that the tensor fields $\phi_{\und{\mu}}(x)$ are space-like.  This is crucial for the unitarity of the emergent field theory. Together with the compactness of the fiber $S^2$ (which is essential for UV finiteness), that is the reason why the present covariant quantum spacetimes provide good backgrounds for the matrix model, in contrast to many other backgrounds proposed in the literature.

The commutation relations \eqref{minimal-cov-algebra} reduce in the semi-classical regime to Poisson brackets\footnote{There are extra terms in $\theta^{\mu\nu}$ for $n\neq 0$ reflecting  nontrivial Chern numbers \eqref{chern-c1} for the internal $S^2$, which characterize the different doubleton representations $\cH_n$.},
\begin{align}
\{t^\mu,x^\nu\} &=  \frac{x_4}r \,\eta^{\mu\nu}\nn\\
 \{x^\mu,x^\nu\} &= - \frac{r^3}{x_4}(t^\mu x^\nu - t^\nu x^\mu)\ =: \theta^{\mu\nu} \ = \
- r^4 \{t^\mu,t^\nu\} \ .
  \label{XT-CR-semi}
\end{align}
The frame and metric are then obtained from \eqref{S-kin-phi-general} as 
\begin{align}
    e^{\dot\a\mu} &= \frac{x^4}{r} \eta^{\dot\a\mu}, \qquad 
    G_{\mu\nu} = \ r^{-4}\sinh(\t)\,\eta_{\mu\nu}  ,
    \nn\\
    \rho^2 &= \rho_M \det(e^{\dot\a\mu})
     = \frac {1}{2\pi^2 r^4}\sinh^3(\tau), 
     \qquad \rho_M = \frac{1}{2\pi^2 r^4 \sinh(\tau)}
     \label{metric-explicit-cov-Cartes}
\end{align}
in Cartesian coordinates $x^\mu$. Here 
$\tau$ is a convenient time parameter $\tau$ on $\cM^{3,1}$ defined by
\begin{align}
    x_4 = r\sinh(\tau) \ 
\end{align}
and $\Omega = \rho_M d^4 x$ is the symplectic volume form reduced to $\cM^{3,1}$.
Note that this frame is classical i.e. without $\hs$ components, and compatible with local $SO(3)$ rotations.
The geometry can be recognized as a (double cover of) a $k=-1$ FLRW spacetime with Big Bounce at $x_4 =0$ \cite{Sperling:2019xar}, 
\harold{with metric and cosmic scale function
\begin{align}
\label{a(t)-explicit}
 ds^2 = G_{\mu\nu} dx^\mu dx^\nu = -dt^2 + a(t)^2 d\Sigma^2, \qquad
    a(t) \sim r^{-1} e^{\frac 32\tau} \ \ \propto \  t 
\end{align}
for late times $ t\to\infty$.
}
A schematic picture of the spacetime and its bundle structure is shown in figure \ref{fig:bundle}.
\begin{figure*}[t!]
    \centering
    \includegraphics[scale=0.5]{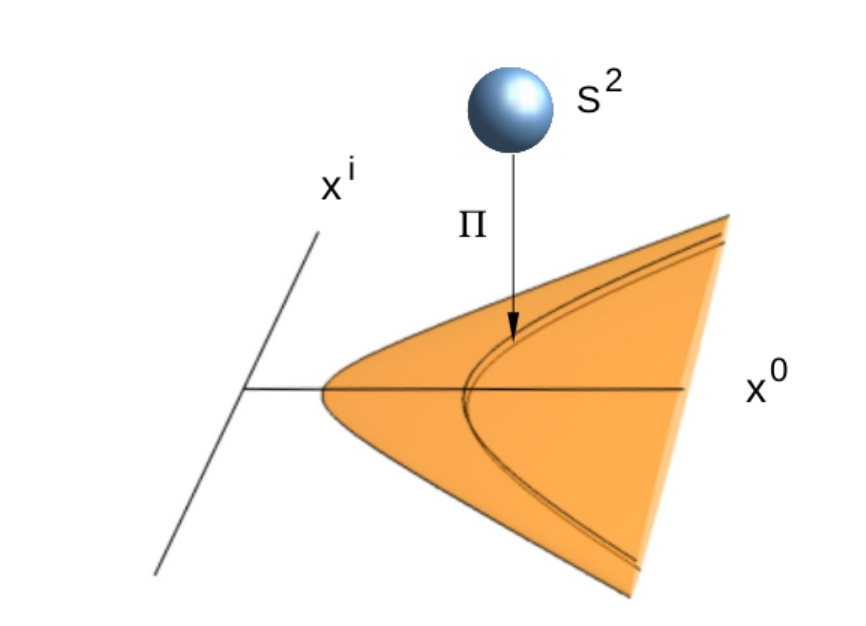}
    \caption{Sketch of the bundle space $\cM^{3,1}\times S^2$. The black lines indicate space-like $H^3$.}
    \label{fig:bundle}
\end{figure*} 

\harold{A remark on the dimensions is in order: since the matrices $\TT^{\dot\a}$ are considered as dimensionless (having absorbed a coupling constant $g$), the parameter $r$ is also dimensionless a priori. It specifies the quantization scale of $X^0$ in the Cartesian coordinates $x^\mu \sim X^\mu$.}

\paragraph{Rigid $k=0$ spacetime.}

A very similar background leads to a FLRW cosmology with $k=0$. It is given by the same doubleton representations of $SO(4,2)$, but with modified matrix configuration $T^\mu$ and spacetime generators $Y^\mu$ satisfying the $E(3)$-covariant relations
\begin{align} 
\label{eq:commut_rel_TY}
[T^i,T^j] &= 0 , \quad  [T^0,T^i] = -\frac i r T^i, \quad
[T^\mu,Y^\nu] = i \frac{Y^0}{r} \eta^{\mu\nu} \nn\\
[Y^0,Y^i] &= i r^3 T^i, \quad 
[Y^i,Y^j] = \frac{-ir^3}{Y^0}(T^i Y^j - T^j Y^i)
\end{align}
(for $n=0$), as well as the constraints
\begin{align} 
T^i T_i &= \frac{1}{r^4} {Y^0}^2, \qquad
T^\mu Y_\mu  + Y^\mu T_\mu =0 \ .
\label{eq:TY_constraints}
\end{align}
More details can be found in \cite{Gass:2025bqr}, and an explicit realization on $\cH = L^2(\R^3)$ is given in \cite{Ho:2025htr}. To be specific, we will mostly focus on the $k=-1$ case in the following.

\paragraph{Emergent gauge theory.}

On covariant quantum spacetime, the action for
 tangential and transversal fluctuations 
\begin{align}
\cA^{a} = 
\begin{pmatrix}
    \cA^{\dot\a} \\ \cA^i
\end{pmatrix} = \begin{pmatrix}
   e^{\dot\a\mu} A_\mu \  \\  \phi^i
\end{pmatrix} 
\label{A-tang-trans-M31}
\end{align}
(which are in general $\hs$-valued!) 
 takes again the form of a $\cN=4$ SYM gauge theory
\begin{align}
    S_{\rm IKKT}[A,\phi] 
&= \int\limits_{\cM^{3,1}} \frac{d^{4} x}{(2\pi)^2}\,\sqrt{|G|}
\Big[- \frac{1}{g_\YM^2} G^{\mu\mu'}\, G^{\nu\nu'}\,F_{\mu\nu}\,F_{\mu'\nu'}
\, - \frac{1}{g_\YM^2} G_{\mu\mu'}\, G_{\nu\nu'}\,B^{\mu\nu}\,B^{\mu'\nu'} \nn\\
& \qquad
 - 2 G^{\mu\nu}\, D_\mu\phi^i D_\nu \phi_i
 + {g_\YM^2} [\phi^i,\phi^j][\phi_i,\phi_j] \Big]_0 
\label{action-YM-scalars-covar}
\end{align}
with $\hs$-valued \harold{fields and} gauge transformations.
Here $B^{\mu\nu} = -\frac{1}{r^2x_4^2}\theta^{\mu\nu}$ can be interpreted as background flux.
The effective coupling is again determined by the dilaton as
\begin{align}
    g^2_\YM = \frac{1}{\rho^2} \ 
    \label{YM-coupling-covar}
\end{align}
which becomes weak at late times for the present rigid background. The time-dependence of the dilaton changes for more general time-dependent backgrounds, as discussed below.
This YM form of the action however hides the geometric nature of the theory, since $\cC^1$--valued gauge fields 
change the background and the effective metric $G_{\mu\nu}$.

A caveat should be noted here: We dropped the linear fluctuations, pretending that the background is a solution. This is true only upon adding a mass term $\mu^2 \tr(\TT^a\TT_a)$ to the action, which is fine classically but not allowed upon quantization since it spoils SUSY. The proper way to proceed is to generalize the matrix background as follows:

\subsubsection{Dynamical covariant cosmological spacetime}
\label{sec:dynamical-covar}

The above rigid covariant spacetime is a classical solution of the model in the presence of a mass term. 
However, an obvious generalization gives the most general 
$SO(3,1)$-covariant
matrix background \eqref{covar-BG}, and provides a solution of the proper IKKT model. 
Since $X_4$ is invariant under $SO(3,1)$, 
the most general covariant background is \cite{Battista:2023glw}
\begin{align}
    \TT^{\dot\a} = \a(X_4) T^{\dot\a} + \beta(X_4) X^{\dot\a} \qquad (+ h.c.) .
\end{align}
Moreover, one can eliminate either $\a(X_4)$ or $\b(X_4)$ using a gauge transformation, and we will choose $\b=0$.
We therefore consider the following matrix background
\begin{align}
\label{generic-covar-background}
      \TT^{\dot\a} = \a(X_4) T^{\dot\a}  \ \sim  \  \a(x_4) t^{\dot\a} \ .
\end{align}
The uniqueness statement implies that this ansatz 
{\em is guaranteed to provide a solution at the quantum level}. 
This ansatz indeed allows to find solutions of the classical IKKT model without mass term, although with problematic behavior \cite{Manta:2025tcl}. Physically reasonable solutions are obtained upon 
 {\em taking into account the 1-loop vacuum energy} in the presence of compact extra dimensions, as discussed around  \eqref{tr-K-1loop} in section \ref{sec:fuzzy-extra-dim}. 
\harold{Explicit computations  \cite{Manta-new} lead to\footnote{This requires some assumptions on the structure of $\cK$, as well as various approximations which are not fully under control.} $\a \sim e^{-\frac 14\tau}$,
which describes an expanding  universe with $a(t) \propto t$ similar to the rigid case, but with slowly increasing dilaton $\rho \propto a(t)^{\frac 45}$ and hence decreasing Yang-Mills coupling $g_\YM$,
consistent with the weak coupling assumption. 
The fuzzy extra dimensions $\cK$ are also stabilized, with a large UV/IR hierarchy
between the UV scale set by $\cK$ and the IR scale set by cosmological spacetime $\cM^{3,1}$.} A detailed discussion will be given elsewhere.  
Analogous statements should apply to the $k=0$ background \cite{Gass:2025bqr}, but its evolution has not yet been worked out.

To understand the fluctuations and the physics on these backgrounds, 
it is necessary to find appropriate local normal coordinates for spacetime, as discussed in section \ref{sec:def-BG-LNC}.

\section{Time evolution, Hilbert space,  and unitarity}

One frequently asked question is how time evolution and
a physical Hilbert space of states can arise in the IKKT model, which -- unlike its $0+1$ dimensional sibling known as BFSS model \cite{deWit:1988wri,Banks:1996vh} -- has no intrinsic time.
We can address
this question on nontrivial backgrounds\footnote{This shows the fundamental difference to holographic approaches, where everything emerges from quantum effects, often in  the trivial vacuum.}. 
Then the matrix fluctuations correspond to (off-shell, a priori) fields on spacetime governed by the matrix d'Alembertian $\Box = [T^{\dot\a},[T_{\dot\a},\cdot]] \sim \rho^2 \Box_G$ with a Lorentzian metric $G$, and the quantization \eqref{correlators-Mink} of the matrix model corresponds to the path integral quantization of QFT.  Hence the Hilbert space and the causal structure of spacetime can be identified from the space of on-shell modes, and their details depend on the type of background.

\subsection{Time evolution}
\label{sec:time-evol}

For any nontrivial background, the time-like frame vector field  \eqref{frame-1}
\begin{align}
     e^0 = -i[\TT^0,\cdot] \sim \{\TT^0,\cdot\}
\end{align}
  defines a unitary 1-parameter evolution operator
\begin{align}
 U(t) = e^{-it[\TT^0,\cdot]}
\end{align}
on the space of matrices.
Since $e^0$ is a canonical vector field, it respects the natural (Hilbert-Schmidt) inner product on $\Mat(\cH) \sim \cC(\cM)$, 
\harold{reflecting the divergence constraint \eqref{div-free} for the frame.}
Therefore
$U(t)$ defines a unitary one-parameter evolution on the space of modes. \harold{This evolution is time-like because $\Box = \eta_{\dot\a\dot \b}e^{\dot\a} e^{\dot\beta} = \rho^{2} \Box_G$  for the effective metric $G = \rho^{-2} \eta_{\dot\a\dot \b}e^{\dot\a} e^{\dot\beta}$ 
in the semi-classical regime \eqref{eff-metric-def}, \eqref{Box-metric-Laplacian}, and related to a physical time evolution by the dilaton $\rho$.
For covariant quantum spacetime, 
$e^0 = \{t^0,\cdot\} = -\sinh(\tau)\frac{\del}{\partial x^0}$
is a time-like vector field, but it is {\em not} proportional to the global cosmic time-like flow defined by the FLRW metric. Similarly
on generic deformed backgrounds, $e^0|_p = -\frac{\del}{\del\tilde y^0}$ (in local normal coordinates for $\gamma^{\mu\nu}$, see section \ref{sec:def-BG-LNC})
is related to {\em some} physical time flow via the dilaton. }
Hence even on a background corresponding to a black hole (yet to be found), the evolution $U(t)$ would still be well-defined and unitary, {\harold{which is certainly remarkable}.

An arrow of time is  encoded in the $i\varepsilon$ regularization \eqref{infinites-Wick}, \eqref{S-mink-reg-mass} of the action, which induces the Feynman $i\varepsilon$ prescription for 2-point functions $\langle \cA(x) \cA(y)\rangle$. These are hence time-ordered, incorporating a decomposition into positive and negative energy modes. This is remarkable, because the background describes an expanding FLRW universe with a Big Bounce \cite{Karczmarek:2022ejn,Battista:2022hqn}.

In particular, 
this discussion strongly suggests that the time-like matrix $\TT^0$ should not be interpreted as time as sometimes advocated \cite{Nishimura:2019qal,Brahma:2021tkh}, but rather as generator of time evolution.
Nevertheless, $\TT^0 \sim {\bf t}^0$ is an interesting function on the underlying symplectic space $\cM$, which for covariant quantum spacetime is a 6-dimensional bundle space.

\subsection{Physical Hilbert space and $\hs$ modes}

Now consider rigid covariant quantum spacetime \eqref{covar-BG} as background, which has many physically desirable properties. 
Its intrinsic higher-spin modes lead however to technical complications, which will be briefly discussed.

The matrix background defines the matrix d'Alembertian $\Box \sim \rho^2 \Box_G$ which governs the free (quadratic) action 
for the fluctuations, which is essentially $\hs$-valued $\cN=4$ 
SYM \eqref{action-YM-scalars-covar}. 
More interesting physics will arise
for product backgrounds $\cM^{3,1} \times \cK$ as discussed below, but for the present purpose it suffices to consider the free action; we always assume or require that these 
fluctuations are weakly coupled. One can then
define the Hilbert space  of physical modes $\cA$ as (properly gauge-fixed)  space of on-shell modes modulo gauge invariance \cite{Kugo:1979gm}
\begin{align}
 \cH_{\rm phys} = \{\mbox{gauge-fixed on-shell modes}\}/_{\{\mbox{pure gauge modes}\}  } 
 \label{H-phys}
\end{align}
This is worked out in detail in \cite{Steinacker:2019awe},
and shown to be a Hilbert space with positive definite norm, without  ghosts. The result is not trivial, because the indices of tangential fluctuations $\cA^a$ are contracted with a Minkowski metric, {\em and} all fields are $\hs$ valued as discussed below. Nevertheless, the Yang-Mills structure of the action together with the constraints \eqref{TT-XX-relation-2} of covariant quantum spacetime ensures that no negative norm states or null states survive in $\cH_{\rm phys}$. 
A detailed exposition of this  structure can be found in \cite{Steinacker:2019awe,Steinacker:2024unq}, and we refer to \cite{Steinacker:2026qzk} for a discussion of the tangential modes related to gravity.

\paragraph{$\hs$ structure of modes.}

To gain some insights, consider the simpler case of transversal fluctuations
 $T^i \equiv \phi \in \cC$  \eqref{A-tang-trans-M31}, decomposed into $\hs$ modes 
\begin{align}
    \Mat(\cH) =\cC = \cC^0 \oplus \cC^1 \oplus \ldots \oplus ... \quad \ .
  \label{EndH-Cs-decomposition}
\end{align} 
In the semi-classical limit, the  $\cC^s$  are modules over $\cC^0$, which encode spin $s$ fields on $\cM^{3,1}$. Indeed, 
they can be organized as \eqref{hs-decomp-spacelike} 
\begin{align}
\cC^s \ni \Phi^{(s)} = \phi_{\mu_1 ... \mu_s}(x) u^{\mu_1} ... u^{\mu_s} , 
 \qquad \phi_{\mu_1 ... \mu_s} x^{\mu_i} = 0
 \label{Cs-explicit}
\end{align}
 in terms of space-like symmetric tensor fields on $\cM^{3,1}$.
 They comprise $2s+1$ degrees of freedom on $\cM^{3,1}$, but local Lorentz invariance is not manifest. This is recovered in an alternative  characterization of $\cC^s$ as space of traceless totally symmetric divergence-free rank $s$ tensor fields on $\cM^{3,1}$, via the bijective correspondence \cite{Steinacker:2024unq,Steinacker:2019fcb}
\begin{align}
\Phi^{(s)} \ \leftrightarrow \
  \tilde\phi^{\mu_1 \ldots  \mu_s}(x) 
  =[\{ ... \{\phi^{(s)},x^{\mu_1}\}, ..., x^{\mu_s}\}]_0  \ ,\quad
\del_\mu(\rho_M \tilde\phi^{\mu_1 \ldots  \mu_s}) = 0
 \label{psi-iso-2-M31}
\end{align}
where $[\cdot]_0$ denotes the projection to $\cC^0$. These are no longer space-like, as illustrated by the frame $e^{\dot\a} = \{\TT^{\dot\a},x^\mu\}$. 
Decomposing the modes $\phi\in\cC^s$  further into $SO(3,1)$ irreps by imposing an on-shell condition  $\Box \sim 0$  leads to a nonstandard organization because $SO(3,1)$ is the group of space-like isometries, rather than the local Lorentz group\footnote{The background does not admit local Lorentz boosts. Local Lorentz invariance is not manifest, but are recovered as part of the volume-preserving diffeos implemented by gauge transformations \eqref{gauge-VF-M}.}. These irreps encode massless divergence-free traceless spin $s$ tensors in radiaton gauge with 2 degrees of freedom, as well as their derivative descendants; we refer to \cite{Steinacker:2019awe,Steinacker:2024unq} for  details.
It is expected that most of these ''massless'' irreps recombine into massive modes through quantum effects in the matrix model, as discussed in section \ref{sec:induced-masses}.

\subsubsection{Deformed covariant spacetime and local normal coordinates}
\label{sec:def-BG-LNC}

We have seen that the matrix background defines in the semi-classical regime a frame. This frame
is in general $\hs$ valued\footnote{This means that the frame is not constant  along the internal $S^2$ fiber.}, but it is always divergence-free \cite{Steinacker:2024unq,Fredenhagen:2021bnw}
\begin{align}
\label{div-free}
   e^{{\dot\a}\mu} = \{\TT^{\dot\a},x^\mu\}  \quad \in \ \cC , \qquad  \del_\mu(\rho_M e^{{\dot\a}\mu}) = 0 = \nabla_\mu^{(G)} (\rho^{-2} e^{\dot\a\mu})
\end{align}
where $\nabla^{(G)}$ denotes the Levi-Civita connection w.r.t. the effective metric $G_{\mu\nu}$ \eqref{eff-metric-def}.
This is  a consequence of the Jacobi identity. 
Such frames are typically not integrable, and it is not evident how to separate spacetime and $\hs$ modes, i.e. base and fiber in the 6-dimensional manifold $\cM$.
That question is addressed in the following.

This question is easy to answer for 
a special class of frames which respect a subalgebra\footnote{this might be interpreted in terms of the frame being constant along a modified bundle projection.} $\tilde\cC^0$ interpreted as functions on (deformed) spacetime $\tilde \cM^{3,1}$, i.e. $e^{\dot\a}[\tilde\cC^0] \subset \tilde\cC^0$. 
Then frame and
torsion tensor \eqref{torsion} are classical i.e. $\tilde\cC^0$ valued, and describe a 3+1 dimensional spacetime.
We denote such backgrounds as {\em pure backgrounds}.
The undeformed background satisfies this condition, and some generalizations are known \cite{Steinacker:2024unq}. However, that property is not generic.

More generally, we focus on backgrounds which are sufficiently mild deformations   of \eqref{generic-covar-background},
\begin{align}
\label{background-deformation}
    \TT^{\dot\a} = \a T^{\dot\a} + \cA^{\dot\a}
\end{align}
 and consider $\cA^{\dot\a} \in \cC$ and the frame $e^{{\dot\a}\mu} = \{\TT^{\dot\a},x^\mu\} \in \cC$ as $\hs$-valued fields on undeformed spacetime $\cM^{3,1}$.
Scalar fields and $\hs$ modes are defined with respect to the undeformed fiber. \harold{In particular, $\cA^{\dot\a} \in \cC^0$ describe Yang-Mills gauge fields,  $\cA^{\dot\a} \in \cC^1$ describes geometric deformations i.e. the gravity sector of the theory, while  $\cA^{\dot\a} \in \cC^s$ for $s\geq 2$ describes higher-spin gauge fields.}

In this setting, one can  always find {\bf local normal coordinates} (LNC) around any point $p\in \cM^{3,1}$, 
such that the frame is classical (i.e. not $\hs$ valued), and the first derivative of the metric vanishes at $p$, as in general relativity. 
In a sufficiently small neighborhood, any $\hs$ components of frame, metric and the transition functions are then negligible, since they drop out at the linearized level in the action due to averaging over $S^2$. Therefore the local physics on the background is governed by a classical (pseudo-) Riemannian metric and frame. This makes the framework a potentially viable basis for physics in 3+1 dimensions.

The precise statement for LNC's is as follows \cite{Steinacker:2024unq,Steinacker:2026qzk}:
for any point $p\in\cM^{3,1} \subset \R^{3,1}$ and sufficiently mild deformation \eqref{background-deformation} of the background,
one can always choose  functions 
\begin{align}
    \tilde y^\mu: \ \ \cM \to \R^{3,1}
   , \qquad \mu = 0,...,3 
\end{align}
on $\cM$ which vanish on the $S^2$ fiber over $p$, such that
$\tilde e^{\dot\a\mu} := \{\TT^{\dot\a},\tilde y^\mu\}$ satisfies 
\begin{align}
   \label{LNC-def}
\tilde e^{\dot\a\nu}|_p &= \eta^{\dot\a\nu},  \nn\\
 \del_\rho\gamma^{\mu\nu}|_p &= 0, \qquad \gamma^{\mu\nu} := \eta_{\dot\a\dot\b} \tilde  e^{\dot\a \mu}\tilde e^{\dot\b\nu} \ .
\end{align}
The same statement holds for $G^{\mu\nu}$ replacing $\gamma^{\mu\nu}$ (but not for both).
Such $\tilde y^\mu$ are denoted as local normal coordinates. They
generate a ''local'' algebra $\tilde \cC^0$ interpreted as functions on (a local patch of) deformed spacetime $\tilde\cM^{3,1}$.
More technically,
they define a deformed bundle projection $\cM \to \tilde\cM^{3,1}$.
Since the $S^2$ fiber is undeformed at $p$, 
 the global averaging map $[\cdot]_0: \cC \to \cC^0$ is compatible with the deformed one at $p$, hence the $\hs$ modes are undeformed at $p$.

By construction, 
this local algebra defines local scalar fields $\tilde\phi(\tilde y)\in\tilde\cC^0$, such that $e^{\dot\a}[\tilde\phi]$ has no $\hs$ components at $p$.
In the vicinity of $p$, these
are the lowest modes among a tower of $\hs$ modes, just like on rigid covariant quantum spacetime.  
The frame $\tilde e^{\dot\a\mu}$ and the metric have arbitrarily small $\hs$ components in a small neighborhood $U$ of $p$, and so do 
the transition functions $\tilde y^\mu(\tilde z)$ for different LNC in nearby patches.  We can therefore
 consistently use the undeformed base space $\cM^{3,1}$, find LNC near any points $p\in\cM^{3,1}$, and
equip $\cM^{3,1}$ with a classical frame and metric in LNC around any point $p$ by simply dropping their $\hs$ components. 
This allows to compute the length of curves and the distance of different points, leading effectively to a classical (pseudo-) Riemannian metric
on $\cM^{3,1}$. That procedure is automatically implemented by the matrix model action\footnote{Therefore it should be consistent with Connes spectral distance for Eucliden covariant quantum spaces.} for the lowest fluctuation modes, due to the averaging  $[\cdot]_0$.

However, the $\hs$ components of torsion and curvature are not eliminated in LNC, and they have physical implications in the action which will be briefly discussed in section \ref{sec:mod-gravity}.

Local normal coordinates are worked out in the simple case of dynamical covariant cosmological spacetimes \cite{Battista:2023glw,Gass:2025bqr}, verifying the above picture.

\subsection{Gauge transformations and volume-preserving diffeos}
\label{sec:diffeos}

Consider gauge transformations $\d_\L\TT^{\dot\a} = \{\L, \TT^{\dot\a} \}$ generated by $ \L  \in \cC^1$.
Then scalar fields $\phi\in\cC^0$ transform 
under volume-preserving diffeos:
\begin{align}
 \d_\L\phi &= \{\L, \phi \} 
 = \xi^\mu \del_\mu\phi ,  \qquad 
 \xi^\mu = \{\L,x^\mu\}\
 \label{gaugetrafo-1-covar}
\end{align}
since
\harold{$0 = \del_\mu(\rho_M \{\L,x^\mu\}) \equiv \nabla_\mu^{(G)}(\rho^2\xi^\mu)$ where $\rho_M$ is the symplectic density \eqref{metric-explicit-cov-Cartes} and $\nabla^{(G)}$ is the Levi-Civita connection w.r.t. the effective metric $G_{\mu\nu}$.
This follows from the general properties of Poisson structures or canonical vector fields, and/or using an adapted derivative $\eth_\mu$ on covariant quantum spacetime \cite{Steinacker:2024unq}.}
The frame (and similar tensors such as the torsion) also transforms tensorially, since  
\begin{align}
 \d_\L e^{{\dot\a}\mu} &\sim   \{\L,e^{{\dot\a}\mu}\} - \{\TT^{\dot\a},\xi^\mu\} 
 =  \xi^\rho \del_\rho e^{{\dot\a}\mu} -  e^{{\dot\a}\rho} \del_\rho \xi^\mu 
 = \cL_\xi e^{{\dot\a}\mu} 
 \label{gauge-VF-M}
\end{align}
using the Jacobi identity. 
Any volume-preserving vector field $\xi^\mu$ on $\cM^{3,1}$ can be obtained in this way for suitable $\L\in\cC^1$  \cite{Steinacker:2024unq}.
Therefore the resulting physics is invariant under (a $\hs$ extension of) volume-preserving diffeos, which underlies the no-ghost result stated above. In this way, local Lorentz invariance for the classical, tensorial sector of the theory is ensured\footnote{We will see that a non-local sector of gravity does actually break Lorentz invariance.}.

\section{Compact extra dimensions}
\label{sec:fuzzy-extra-dim}

The physics resulting from the matrix model on covariant quantum spacetime itself would not be very interesting. To obtain interesting physics, richer backgrounds are required where the 6 transversal scalar fields acquire structure, similar to string theory compactifications. There is indeed a prescription for compactifying the IKKT model on (non)commutative tori by imposing $\TT^i \cong \TT^i + c^i \one$ up to gauge transformations \cite{Connes:1997cr}, leading  to features familiar from string theory.
However, this prescription  requires an infinite-dimensional Hilbert space, with infinitely many degrees of freedom (dof) per spacetime volume; then the nonperturbative power of the matrix model is lost.

We therefore {\em do not} adopt such a procedure, but use a different mechanism to achieve our goal, via {\bf fuzzy extra dimensions}. Despite the name this is an entirely different ansatz, which entails only finitely many dof per volume and thereby maintains the non-perturbative power of the matrix theory. This is achieved by giving a  nontrivial background to
{\em all} matrices, typically with a product structure $\cM \times \cK$
\begin{align}
    \label{eq:background-2}
\langle {\bf T}^{A} \rangle =
\binom{\TT^{\dot\a}\otimes \one_\cK}{\one_{\cM}\otimes \cK^{i}}
   \,,\qquad  \dot \a= 0,1, 2, 3\,,\qquad {i}=4,\ldots,9\ , 
\end{align}
which can be viewed as product space embedded in target space $\R^{9,1}$.
Here $\cK$ will typically  be some compact fuzzy (i.e. quantized symplectic) space, leading a truncated tower of Kaluza-Klein (KK)  modes 
\begin{align}
    \Box_6 \Upsilon_\L = m^2_\L \Upsilon_\L , \qquad  m_{\L}^2 = m_\cK^2 \,\mu^2_{\L}\,
\end{align}
with scale $m_\cK^2$ set by the internal matrix Laplacian $\Box_6 = [\cK^i,[\cK_i,\cdot]]$. This can be interpreted as Higgs effect due to SSB via scalar fields in an underlying $U(k)$ gauge theory with background $\TT^{\dot\a} \otimes \one_k$
 \cite{Aschieri:2006uw}.
Two classical mechanisms to stabilize  $\cK$ are known:
\begin{enumerate}
    \item 
    by imposing an $R$ charge i.e. internal rotation of $\cK$ along $\cM^{3,1}$ \cite{Manta:2025tcl,Steinacker:2014eua,Iso:2015mva}, through an ansatz $\cK_i^\pm = e^{i\varphi(\tau)} T_i^\pm$ for 3 complex transversal matrices. Then the kinetic contribution balances the quartic potential of $\cK$, leading to an effective potential for $T_i$ with stable nontrivial minimum. 
    The generation of radiation is avoided by imposing the condition 
\begin{align}
\label{K-vanish}
    \sum_{i}  [\cK^{i +},\cK_{i}^-] 
    \stackrel{!}{=} 0
\end{align} 
which ensures that gauge current $J_\mu = [\TT_{i},D_\mu \TT^{i}]$ vanishes.
    This mechanism is appealing because a large $R$ charge stabilizes a large hierarchy between the UV scale on $\cK$ and the IR scale of spacetime.
    
    \item 
    cubic terms such as in the polarized IKKT model \cite{Bonelli:2002mb}
    can stabilize $\cK$, although maximal SUSY appears to imply the ''wrong'' mass term for spacetime $\cM^{3,1}$.
    Nevertheless, this may be a promising direction to explore; cf. \cite{Hartnoll:2024csr,Komatsu:2024bop,Komatsu:2024ydh,Ciceri:2025wpb} for a holographic interpretation of this model.

\end{enumerate}

On the other hand, quantum effects are expected to (help) stabilize $\cK$, which remains to be elaborated in detail, cf. \cite{Manta:2024vol,Steinacker:2024huv,Manta-new}.
The stabilization mechanism via $R$ charge may have interesting ramifications for the low-energy gauge theory, notably in regard to chirality and CP violation, which remain to be understood. There are interesting examples of $\cK$ where the lowest KK modes have vanishing $R$ charge and are chiral, being localized at (self-)intersections of $\cK$ at the origin of $\R^6$ \cite{Steinacker:2014eua,Sperling:2018hys}. These seem to have the low-energy properties required to obtain an effectively Lorentz-invariant low-energy field theory.

\harold{The intrinsic structure of $\cK$ determines not only the effective nonabelian gauge theory and its Kaluza-Klein spectrum, but also the gravitational physics as discussed below. At first sight, this amounts to a sort of ''mini-landscape'' problem similar as in string theory. However, the present problem is much more tractable, because  $\cK$ is governed by the (effective) matrix model action, leading to an optimization problem for matrix configurations. The same of course applies to the full matrix background $\cM \times \cK$, which from a string theory point of view corresponds to determining branes embedded in un-compactified target space. 
There might still be many possible such (meta-) stable matrix vacua or brane configurations, and determining them is an important but accessible problem in the present framework.}

\section{One-loop effective action and gravity}
\label{sec:one-loop-gravity}

There is a universal background-independent formula for the 1-loop effective action of the IKKT model.
Taking into account the fermions and the gauge-fixing ghosts in the IKKT model,
the one-loop effective action defined by the Gaussian integral around the background 
\begin{align}
 \int\limits_{\rm 1\, loop} dT d\Psi d\bar c dc\,
 e^{iS[T,\Psi,c]}
 = e^{i (S_0[T] + \Gamma_{\!\rm{1 loop}}[T]) } \
  = e^{i \Gamma_{\rm eff}[T]} \ 
 \end{align}
takes the explicit form 
\cite{Ishibashi:1996xs,Chepelev:1997av,Blaschke:2011qu,Steinacker:2024unq}
\begin{align}
\Gamma_{\!\textrm{1loop}}[\TT]\!
&= \frac i2 \Tr \Big(\!\log(\Box \! -\! i\varepsilon \! - \!\Sigma^{(\cA)}_{ab}[ \cF^{ab},\cdot])
- \!\frac 12 \log(\Box \! - \!i\varepsilon \! - \!\Sigma^{(\psi)}_{ab}[ \cF^{ab},\cdot] )
- \! 2 \log (\Box \! - \! i\varepsilon )\Big)   \nn\\
  &= \frac i2 \Tr \Big(-\frac 14 \big((\Box -i\varepsilon )^{-1}\Sigma^{(\cA)}_{ab} [ \cF^{ab},\cdot] \big)^4
  +\frac 18 \big((\Box -i\varepsilon )^{-1}\Sigma^{(\psi)}_{ab} [ \cF^{ab},\cdot]\big)^4 + ...\nn\\
  &\sim  \frac {3i}4 
  \int\limits_{\cM\times\cM}\frac{\Omega_x \Omega_y}{(2\pi)^m}
  \sum_\L \frac{V_4[ \d\cF({ x},{ y})] }{(|x-y|^2+ m_\L^2 -i\varepsilon)^4} + ...
\label{Gamma-IKKT}
\end{align}
 up to higher-order terms in $\cF_{ab} = i[\TT_a,\TT_b]$. The last expression holds for backgrounds $\cM \times \cK$ where $\Mat(\cH_\cM)\sim \cC(\cM)$ can be interpreted as quantized space of functions on  $\cM$ \eqref{semi-classical}, summing over KK modes on $\cK$. Here $\d \cF(x,y) = \cF(x) - \cF(y)$
 and 
\begin{align}
 V_4[ \d\cF] = -4\tr (\d\cF^4) + (\tr \d\cF^2)^2 \
 \label{S-4-short}
\end{align}
where $\tr$ denotes the trace over the $ab$ indices  with $\eta_{ab}$.
Several features should be observed:

\begin{itemize}
    \item 
The first three terms in the expansion of the $\log$ cancel identically due to maximal SUSY, for any background $\TT$.  Moreover, $\Gamma_{\!\textrm{1loop}}[\TT]$ is invariant under rescaling $\TT \to c\TT$.

\item The 1-loop action vanishes for commuting backgrounds and for backgrounds where $\cF^{ab}$ is central, such as Moyal-Weyl.

\item On a nonabelian Moyal-Weyl backgrounds in the Coulomb branch, this effective action is consistent with 
an expansion of the DBI action for a
brane with the 9+1 dimensional  $AdS^5 \times S^5$ target space  metric \cite{Steinacker:2024unq}. This exhibits the  relation with IIB string theory, which is however not the point here.

\item 
The last form  in \eqref{Gamma-IKKT} is the   bi-local 1-loop effective action  {\em in position space} on $\cM$, obtained by computing the trace $\Tr_\cM$  over the space of all modes   using 
 string modes $|x\rangle\langle y|$ \cite{Steinacker:2023myp,Steinacker:2024unq}. 
 This is of course non-local on $\cM$, but it is {\em weakly non-local} i.e. the locality induced by string modes in the loops is {\em short-range}, suppressed as $\frac1{|x-y|^8}$. This also reflects the relation with  IIB supergravity in target space $\R^{9,1}$.

\item 
$ V_4[ \d\cF] \leq 0$ for Euclidean $\cF_{ab}$ of rank $\leq 4$.  This should be important to form bounds states in $\cK$.
 
\item For single matrix models, the  Faddeev-Popov determinant is nothing but the
Vandermonde determinant, which leads to a good description of the models via an eigenvalue distribution \cite{forrester2010log}. Hence the one-loop approximation for multi-matrix matrix models should be more powerful than one might naively expect.
    
\end{itemize}

These 1-loop computations are  finite and meaningful for two reasons:
1) {\em maximal SUSY} of the matrix model\footnote{Note that the background is not required to respect SUSY, and of course it should not.}, and 2) {\em finitely many dof} (per volume) of the underlying quantum space.
In models with less or no SUSY, the non-local contributions would be long-range, and entail pathological UV/IR divergences at higher loops in 4 dimension. This singles out the IKKT model 
\harold{among all Yang-Mills matrix models}
as essentially unique candidate for a fundamental physical theory\footnote{This is analogous to superstring theory, which is singled out among other string theories \harold{by modular invariance, which tames UV divergences and prevents runaway IR instabilities}.} \cite{Jack:2001cr,Matusis:2000jf,Steinacker:2024unq}.
Infinitely dof per volume would lead to divergent internal traces (e.g. for the KK modes) in the explicit expressions given below.

\subsection{Vacuum energy: evading the cosmological constant problem}

In general relativity, the Einstein equations 
\begin{align}
\cG_{\mu\nu} + \Lambda g_{\mu\nu} = 8\pi G_N T_{\mu\nu}    
\end{align}
describe the gravitational dynamics of massive objects. However they also lead to the notorious cosmological constant problem, as vacuum energy curves spacetime. 
Vacuum energy is generically large on non-supersymmetric backgrounds, since it includes the zero-point energy due to  quantum fluctuations of all modes below the SUSY breaking scale. This entails a large curvature of spacetime, in conflict with observation. The only way to avoid this clash within GR is to postulate some finely tuned a priori cosmological constant $\Lambda$. The associated unreasonable fine-tuning constitutes the cosmological constant problem.

In the present framework, this problem  does not arise, because the Einstein equations are only valid on intermediate scales.
At one loop, the dominant contribution to 
the vacuum energy arises from $\cK$, given by \cite{Steinacker:2023myp}
\begin{align}
   \Gamma_{vac,\rm loc}^\cK 
   =  \frac{3i}4\Tr\Big(\frac{V_{4,\cK}}{(\Box - i \varepsilon)^4}\Big)
  \sim
  \int \Omega \,e^{-\tau} \Big(\frac{r m_\cK}{\a} \Big)^4 \sum_{\L}
  \frac{V_{4,\L}}{\mu^4_{\L}} 
  \  
   \label{tr-K-1loop}
\end{align}
where $\sum_\L$ indicates the finite sum over KK modes. 
A sum over $\hs$ modes truncated by \eqref{hs-cutoff} is suppressed. This is typically large, set my the UV scale $m_\cK$.
However at very large (cosmic) scales, the geometry of spacetime is determined by the combination of the Yang-Mills action with vacuum energy, while the Einstein-Hilbert action is suppressed and becomes irrelevant. Then
a large vacuum energy \harold{is in fact essential to stabilize asymptotically flat covariant quantum spacetime, as stated in section \ref{sec:dynamical-covar} and detailed in \cite{Manta-new}. 
No fine-tuning of a cosmological constant is needed, but suitable properties of the $V_{4,\L}$ in the KK spectrum are required.
Since $\cK$ is determined dynamically by its 1-loop potential $\Gamma_{vac,\rm loc}^\cK$ combined with the classical actions, the associated vacuum selection is a well-defined problem.}

\subsection{Induced Einstein-Hilbert action}
\label{sec:E-H}

 Although the matrix model defines a dynamical quantum geometry, it does not resemble GR unless quantum effects are included. It turns out that 
  the Einstein-Hilbert action arises in the 1-loop  effective action  on quantum spacetime $\cM^{3,1}$, but {\em only} in the presence of $\cK$. The reason is that the 1-loop  action for $\cM \times \cK$ includes a mixed term from $V_4[ \d\cF]$
which couples the torsions of $\cM^{3,1}$ and $\cK$, which then reduces to the Einstein-Hilbert action on $\cM^{3,1}$.
That term in 
$\Gamma_{\rm 1 loop}$
is \cite{Steinacker:2021yxt,Steinacker:2023myp}
\begin{align}
 \Gamma_{\rm 1 loop}^{\cK-\cM}  
 &= - \frac 12\int\limits_\cM d^4 x  \,\frac{\sqrt{G}}{16\pi G_N}
   \tensor{T}{^\rho_\sigma_{\mu}}\tensor{T}{_{\rho}^{\sigma}_{\nu}} G^{\mu\nu}\
  =  \! \int \! d^{4}x\frac{\sqrt{|G|}}{16 \pi G_N}\,
   \Big(\cR
  + \tfrac 12\tilde T_{\nu} \tilde T_{\mu}  G^{\mu\nu} \!
        - 2 \rho^{-2} \del_\mu\rho\del^\mu\rho
 \Big) 
   \label{Gamma-EH-0}
\end{align}
where $\tilde T_\mu = \rho^{-2}\del_\mu \tilde\rho$ is an axionic vector field defined by the totally antisymmetric torsion \cite{Fredenhagen:2021bnw}, and the Newton constant is determined by the structure of $\cK$ as \cite{Manta:2024vol}
\begin{align}
\frac{1}{G_N}&\equiv
\frac{1}{4\pi\rho^2}\Tr_\cK\Big(\frac{\delta\cF_{ij}\delta\cF^{ij}}{\Box_6}\Big)
\end{align}
up to higher order terms in $\d \cF = [\cF,\cdot]$. \harold{Its scale is set by the KK scale $m_\cK$ inherent in $\cF_{ij}$ and $\Box_6$, but also on large numbers arising from the finite sum over KK and $\hs$ modes; more explicit expressions can be found in \cite{Steinacker:2023myp,Manta-new}. In particular, different $\cK$  lead to different values of the Newton constant.}


The main message is that gravity arises as quantum effect  on quantum spacetime in the IKKT model on suitable backgrounds.
\harold{From the string theory perspective, the present mechanism could be interpreted in terms of branes in a Seiberg-Witten type limit $\a' \to 0$\cite{Seiberg:1999vs}, where the closed string sector decouples. The relevant gravitational metric is then the 3+1-dimensional open string metric  on the brane, rather than the closed string metric in 9+1-dimensional (uncompactified) target space. 
This is the origin of the non-standard properties of the gravity theory described below.
The specific realization of covariant quantum spacetime as brane in string theory is not known, see however \cite{Castelino:1997rv} for a related discussion.}

\paragraph{Propagating gravity modes.}

The detailed physics of the gravitational sector of the theory is considerably richer than GR, and remains to be understood in detail. 
To gain some insight, one can elaborate -- in the locally flat regime around the reference point $\xi^\mu = (\xi^0,0,0,0) \in \cM^{3,1}$  -- 
the gravitational dof. on covariant quantum spacetime following \cite{Steinacker:2026qzk}. 
Consider $\cC^1$ valued perturbations $\TT^{\dot\a} \to  {\bf T}^{\dot\a} = t^{\dot\a} + \cA^{\dot\a}$ organized as plane waves 
\begin{align}
\label{A-general-onep}
    \cA^{\dot\a} = A^{\dot\a \mu}(x) u_\mu, \qquad A^{\dot\a \mu}(x)  = A^{\dot\a \mu} e^{i k x} , \  \  A^{\dot\a 0} = 0 
\end{align}
with on-shell condition\footnote{This applies to the Yang-Mills action and to the 1-loop effective action at quadratic order. However their combination leads to extra modes with modified dispersion relation discussed in section \ref{sec:mod-gravity}.} $\Box  \cA^{\dot\a} = 0 = k_\mu k^\mu$ after gauge-fixing $\del_{\dot\a} A^{\dot\a \mu} = 0 = k_{\dot\a} A^{\dot\a \mu}$. 
Subtracting the 3 residual on-shell pure gauge modes $k^{\dot\a} \L^\mu$ for $\mu = 1,2,3$, we obtain 6 = 4+1 + 1 physical on-shell modes, which might be viewed as 4 modes of a partially massive graviton, a conformal (dilaton) mode and an axion. 
Specifically, for light-like
\begin{align}
\label{k-null-explicit}
    k^{\dot\a} = (k,k,0,0) \ \equiv k^+ 
\end{align}
the 6 physical modes are
\begin{align}
\cH_{phys} = \{A^{2 \mu}, \  A^{3 \mu} \}
\end{align}
for space-like $\mu=1,2,3$, which satisfy the space-like condition
\begin{align}
\label{spacelike-A}
    \xi_{\dot\a} A^{\dot\a \mu} = 0 \ .
\end{align}
We can separate these into 4 transversal modes $A_{(tran)}^{\dot\a \mu} k_\mu = 0$,
and 2 longitudinal modes $A_{(long)}^{\dot\a 1}$ which will lead to extra $\mu=+$ frame modes \eqref{physical-e-modes}.

Recall that $\cA$ is a {\em potential} for the frame.
The effective frame modes are 
obtained as 
\begin{align}
\label{frame-modes}
    \d e^{\dot\a \nu} 
    &= \{A^{\dot\a \mu}u_\mu,x^\nu\}
    \propto  A^{\dot\a \mu} u_\mu k_\sigma (\xi^\sigma u^\nu - \xi^\nu u^\sigma) 
\end{align}
which is divergence-free 
\begin{align}
\label{e-div-free}
\del_\nu \d e^{\dot\a \nu}  
 = 0 
\end{align}
and transversal space-like in $\dot\a$ 
due to the Lorentz gauge condition for $A$ and
\begin{align}
\label{spacelike-e}
    \xi_{\dot\a} \d e^{\dot\a \mu} = 0  = \del_{\dot\a} \d e^{\dot\a \nu} \ .
\end{align}
However they are no longer space-like in $\nu$.
The 6 physical frame modes are given by
\begin{align}
\{\d e^{2 \mu}, \  \d e^{3 \mu} \} \qquad \mbox{for}\ \  \mu = +,2,3 \ ,
\label{physical-e-modes}
\end{align}
which contain components in $\cC^0$ and $\cC^2$.
The classical $\cC^0$ components of $A_{(tran)}^{\dot\a\mu}$ lead to 4 transversal frame modes with $\mu=2,3$, while 
the two $\mu=+$ modes arising from $A_{(long)}^{\dot\a 1}$
are longitudinal $[\d e^{\dot\a \mu}_{(long)}]_0 \propto w^{\dot\a} k^\mu$. The latter lead to diffeo-like helicity $\pm 1$ graviton modes (which are nevertheless physical frame modes), while the former lead to 2 transversal graviton modes and a trace mode, as the gauge group consists of volume-preserving diffeos only. All graviton modes satisfy the harmonic gauge condition $\del_\mu h^{\mu\nu} =0$.
The remaining physical frame mode is captured by the axion, and possibly in the $\cC^2$ components of the graviton.

\subsection{Induced mass terms and massless modes}
\label{sec:induced-masses}

Quantum fluctuations of the background 
lead to induced mass terms\footnote{The 
$\hs$ modes are massless at the classical level because the background frame generates a rank 4 effective metric on $\cM = \C P^{1,2}$. Adding quantum fluctuations to the background
will contribute kinetic energy along the fiber, acting like an effectively 6-dimensional metric on $\cM$, leading to mass terms for the $\hs$ modes. }  $\Tr(\cA^{\dot\a} \cA_{\dot\a})$ in the 1-loop effective action, as shown by explicit computations \cite{Steinacker:2024huv}. This is important on covariant quantum spaces, because it removes the plethora of $\hs$ modes from the low-energy physics, notably the massless $\hs$-valued nonabelian fields. On the other hand it is puzzling, since mass terms for gauge fields are usually incompatible with  gauge transformations
\begin{align}
\delta_\L \cA^{\dot\a} &=  [\cA^{\dot\a},\L] + [\TT^{\dot\a},\L] \ .
\label{gauge-A-phi-NC}
\end{align}
For noncommutative backgrounds however, mass terms {\em can} be compatible with gauge transformations for the following reason:
A quadratic term in the action of the form
\begin{align}
    \Tr (\TT^{\dot\a} + \cA^{\dot\a})(\TT_{\dot\a} + \cA_{\dot\a}) 
    &= {\rm const} + 2\Tr(\TT^{\dot\a} \cA_{\dot\a}) + \Tr(\cA^{\dot\a} \cA_{\dot\a}) 
    \label{quadratic-term-T+A}
\end{align}
is clearly invariant under 
gauge transformations
\begin{align}
    \TT^{\dot\a} + \cA^{\dot\a}  &\to U^{-1} (\TT^{\dot\a} +\cA^{\dot\a}) U .
\end{align} 
We can check explicitly that $\Tr(\cA \cA) + 2 \Tr(\TT \cA)$ is gauge invariant:
\begin{align}
    \delta_\L \Tr(\cA_{\dot\a} \cA^{\dot\a}) &= 2 \Tr(\cA_{\dot\a} [\cA^{\dot\a},\L]) + 2\Tr(\cA_{\dot\a} [\TT^{\dot\a},\L]) 
    = 2 \Tr(\cA_{\dot\a}[\TT^{\dot\a},\L])
\end{align}
which precisely cancels with 
\begin{align}
     \delta_\L 2\Tr(\TT_{\dot\a} \cA^{\dot\a}) &=
     2\Tr(\TT_{\dot\a} [\cA^{\dot\a},\L]) +
     2\Tr(\TT_{\dot\a} [\TT^{\dot\a},\L])
      = 2\Tr(\TT_{\dot\a} [\cA^{\dot\a},\L]) \ .
\end{align}
This mechanism is particularly transparent in the ARS model, where adding a mass term will modify the radius of the vacuum sphere  $\TT^a \to \a\TT^a$, while tangential fluctuations
\harold{$\cA^{\dot\a} = \d \L^{\dot\a}_{\ {\dot\b}}(x) \TT^{\dot\b}$ 
for $x$-dependent $SO(3)$ generators $\d \L^{\dot\a}_{\ {\dot\b}}(x)$}
around the new background give rise to 2 massless modes corresponding to Goldstone bosons of $SO(3)/U(1)$. 
\harold{These are 2 rather than 3 local tangential modes, since rotation modes around the axis $\langle p|\TT^{\dot\a}|p\rangle$ through some point $p\in S^2$ vanish there. As usual in 2D Euclidean Yang-Mills, one is a pure gauge mode, while the other gives rise to gauge-invariant (off-shell) field strength fluctuations. Note that covariance \eqref{covariance-S2} does not mean that all these $x$-dependent modes are pure gauge.
Finally, the radial mode gives rise to a scalar field.}

Analogous considerations apply to 
covariant quantum spacetime in the IKKT model, which in fact requires an induced mass term (or vacuum energy) for stabilization, as discussed above. 
The $\cC^1$-valued deformations $\cA^{\dot\a} = \d \L^{\dot\a}_{\ {\dot\b}}(x) \TT^{\dot\b}$ 
for $x$-dependent $SO(3,1)$ generators $\d \L^{\dot\a}_{\ {\dot\b}}(x)$ can be considered as \harold{6 Goldstone bosons of the global $SO(3,1)$ symmetry\footnote{Now $SO(3,1)$ acts freely on $\cC^1$ over spacetime, hence there is no reduction.}, which guarantees that these are flat directions of the quantum effective potential. 
These are  part of the fluctuation modes in \eqref{A-general-onep} and seem to include the helicity $\pm 1$ gravitons and the axion, but not the 
standard transversal traceless modes of physical gravitons. 
On the other hand, mass terms for those are forbidden as usual
since they transform as
$\d h_{\mu\nu} = \cL_\xi h_{\mu\nu}$ under volume-preserving diffeos.
Further details remain to be elaborated.}

For $\hs$-valued fluctuations $\cA^{\dot\a} \in \cC^s$ with $s > 1$, the protection mechanism via Goldstone modes does not apply to these, since no VEV exists for spin $s>1$.
Therefore these are expected to acquire genuine (large) mass through quantum effects; note that
a mass term $\Tr(\cA_{\dot\a} \cA^{\dot\a})$ is  gauge invariant, since the mixed term $\Tr(\TT_{\dot\a} \cA^{\dot\a})$ vanishes identically. Note that most of the nonabelian YM gauge fields $\cA^{\dot\a} \in \cC^0$ acquire large KK masses via the Higgs mechanism \cite{Aschieri:2006uw}, while the unbroken sector defines the massless YM gauge fields.

\section{Large-scale modifications of gravity \& mirage matter}
\label{sec:mod-gravity}

We have seen that
in the matrix model, the fundamental degrees of freedom are matrices and their fluctuations $\cA$, which play the role of {\em potentials} for the frame \eqref{frame-1}. This leads to important distinctions from the traditional formulation and mechanism of gravity. 
In particular, the semi-classical matrix action is of Yang-Mills type, and {\em cannot} be rewritten as a local action for the frame. This entails some rather unfamiliar IR modifications of gravity.

At one loop, the gravity sector of the effective   geometrical  gauge theory contains (at least) the following three types of terms
\begin{align}
    S_{\rm grav} = S_{\rm EH} + S_{\YM} + S_{\rm vac}
\end{align}
where
$S_{\rm EH}$ is the Einstein-Hilbert term (extended by dilaton and axion) \eqref{Gamma-EH-0}, $S_{\YM}$
is the semi-classical  Yang-Mills type matrix action, and $ S_{\rm vac}$ is the induced vacuum energy.
These can be computed explicitly at weak coupling. 
Let us focus on the YM term 
\begin{align}
\label{S-YM}
    S_{\rm YM} =  -\int
 d^4x \sqrt{G}\rho^{-2} \cF^{\dot\a\dot\b} \cF_{\dot\a\dot\b} \ , \qquad
 \cF_{\dot\a\dot \b} = -\{\TT_{\dot\a},\TT_{\dot\b}\} \ 
\end{align}
which cannot be written as a local function of the frame. To cast the equations of motion into the  form of generalized Einstein equations, we rewrite the variation of this term as 
\begin{align}
  \delta S_{\rm YM} &= - 4\int d^4x \sqrt{G} \rho^{-2}\, \d e^{\dot\alpha\mu}  \, C_{\dot\a\mu} ,   \qquad \d e^{\dot\alpha\mu} = \{\d \TT^{\dot\alpha},x^\mu\}
\end{align}
where the ''anharmonicity tensor'' $C_{\dot\a\mu}$ \cite{Kumar:2023bxg} incorporates the background flux via 
$\{C_{\dot\a\mu},x^\mu\} = - e^{\dot\b}[\cF_{\dot\b\dot \a}]$ up to an ambiguity $C_{\dot\a\mu} \to C_{\dot\a\mu} + \partial_\mu C_{\dot\a} $.
An approximate solution is
\begin{align}
\label{C-explicit-formula-1}
    C_{\dot\a\mu} 
    \ \approx \ -r^{-4}\a^2\widetilde\Box^{-1} \{e^{\dot\b}[\cF_{\dot\b\dot \a}],x^{\mu'}\}\gamma_{\mu\mu'}
\end{align}
where \harold{(see Appendix A.7 in \cite{Steinacker:2026qzk})
\begin{align}
    \widetilde \Box := r^{-4} \a^2 \{x^\mu,\gamma_{\mu\nu}\{x^\nu,\cdot\}\} \sim  \frac 13\a^{2} e^{\t}\Box_{G,c\to\frac 1{\sqrt{3}} c}
\end{align}
}
is a modified d'Alembertian with speed of light reduced by $\frac 1{\sqrt{3}}$, including also $\hs$ contributions which we drop. 
The variation of the YM term can then be cast into the geometric form 
\begin{align}
\label{variation-YM-C}
  \delta S_{\rm YM} &= - 4\int d^4x \rho_M C_{\dot\a\nu} \{\d \TT^{\dot\alpha},x^\nu\} \nn\\
 &\equiv  - \int d^4x\sqrt{G}\rho^{-2}\,
 \big(T_{\mu\nu}[C] - \tfrac{1}{2} G_{\mu\nu}T[C] + B_{\mu\nu}[C]\big)
 e_{{\dot\a}}^{\ \mu} \d e^{{\dot\a}\nu} \ .
\end{align}
Here $T_{\mu\nu}[C]$ is an effective energy-momentum (em) tensor, interpreted as "mirage" matter
since it arises
as a "non-local reflection" of ordinary matter $T_{\mu\nu}$  as shown below. Dropping the antisymmetric term $B_{\mu\nu}[C]$,
we thus obtain modified Einstein equations\footnote{Note that the divergence constraint \eqref{e-div-free}  of the frame is taken care of by the above ambiguity of $C_{\dot\a\mu}$, which is used to make $T_{\mu\nu}[C]$ divergence-free.}
\cite{Steinacker:2026qzk,Kumar:2023bxg}
\begin{align}
  \label{mod-Einstein}
\frac 1{8\pi G_N} \cG_{\mu\nu} 
= T_{\mu\nu} + T_{\mu\nu}[C] \  - G_{\mu\nu} \tilde\Lambda \ .
\end{align}
At the linearized level, the mirage em tensor can be written explicitly using \eqref{C-explicit-formula-1}  as 
\begin{align}
\label{mirage-E-M-metric}
       \d T^{\mu\nu}[C] 
      &\approx  -2 \a^{2} r^{-4}\widetilde\Box^{-1}\Box_G \d \bar G^{\mu\nu } 
\end{align}
where $\d \bar G^{\mu\nu} =  \d G^{\mu\nu} - \frac 12
     G^{\mu\nu}(G_{..}\d G^{..})$
is the trace-reversed metric perturbation. 
This satisfies the harmonic gauge condition $ \nabla^{(G)}_\mu \delta \bar G^{\mu\nu} = 0$ due to the divergence constraint \eqref{div-free}, provided the Weitzenböck torsion of the background is negligible (see  Appendix A.8 of \cite{Steinacker:2026qzk}).
Then the linearized Einstein equation
  $\d \cG^{\mu\nu} 
= 8\pi G_N (T^{\mu\nu} + \d T^{\mu\nu}[C])$ 
takes the form
\begin{align}
  \big(1 - m^2 \widetilde\Box^{-1}\big) \d \cG^{\mu\nu} 
  \approx  8\pi G_N T^{\mu\nu}
  \label{lin-mod-Einstein-4}
\end{align}
in terms of the scale\footnote{This scale is in Cartesian local normal coordinates, see \cite{Steinacker:2026qzk} for a more careful discussion.}
\begin{align}
\label{m-cross-def-1}
m^2 := 32\pi G_N \a^2 r^{-4}\ 
\end{align}
which marks the cross-over between the GR and YM regime, in Cartesian normal coordinates. 
\harold{Measuring this scale with the effective metric gives the (dynamical) cross-over scale $m^2_{\rm cross} = 3 r^{4}\a^{-2}e^{-\tau}$, which is found to satisfy  \cite{Steinacker:2026qzk}
\begin{align}
    \frac{L_{\rm cross}}{L_{\rm UV,G}}
    &\sim g_{\rm YM}  
    \Big(\frac{L_{\rm cosm}}{L_{\rm UV,G}}\Big)^{\frac 23} 
    \frac{L_{\rm UV,G}}{L_{\rm Pl}} \ .
\end{align}
Here $L_{\rm UV,G}$ is the UV scale defined by the quantization of 4-volume, where local QFT must break down. These hierarchies may be large enough to be physically viable; e.g. $L^{-1}_{\rm UV,G} \approx 10^{16}$eV  with $g_{\rm YM} = 1$   leads to  $L_{\rm cross} = 10^{22}m$ being 10 times the galactic scale. However, $L_{\rm cross} = L_{\rm cosm}$ in the weakly coupled regime $g_{\rm YM} \leq 1$ would imply $L^{-1}_{\rm UV,G} \approx 7.5$ TeV, which is no longer viable. Therefore new physics between ths cosmological scale and the Planck scale is bound to arise in the present framework.}


To get some insights into the implications of mirage matter, we rewrite the 
 linearized (modified) Einstein equation \eqref{lin-mod-Einstein-4} as
\begin{align}
     \label{lin-mod-Einstein-5}
\d \cG^{\mu\nu} 
  &\approx  8\pi G_N (T^{\mu\nu} + \delta T^{\mu\nu}[C]) , \qquad  \d T^{\mu\nu}[C] := \frac{m^2}{\widetilde\Box - m^2} T^{\mu\nu} \ .
\end{align}
This \harold{allows to interpret} mirage matter as a non-local reflection of ordinary matter.
A tentative fully non-linear form of these equations was also given in \cite{Steinacker:2026qzk}
\begin{align}
\label{T-C-expression-metric-2}
  \frac{1}{8\pi G_N}\cG^{\mu\nu} \ = \   T^{\mu\nu} + T^{\mu\nu}[C] - G^{\mu\nu} \tilde\Lambda \   
      \ \approx \ \frac{1}{m^2} \widetilde\Box \, T^{\mu\nu}[C] \ .
\end{align}
All this should be taken with a grain of salt, since these quantities are $\hs$ valued.
The action $S_{\rm EH} + S_{\YM}$  including $\hs$ components to quadratic order is worked out in \cite{Steinacker:2026qzk}.

The resulting modifications of GR are not well understood up to now.
 The most obvious and interesting conclusion is that gravity is modified in the IR in a non-local manner, distinct from any conventional modifications of GR. 
For  scales shorter than $m^{-1}$, we recover (linearized) GR, since
$\delta T^{\mu\nu}[C] \approx 0$ for $\widetilde\Box \gg m^2$ \eqref{lin-mod-Einstein-5}. For longer distances, a screening behavior due to mirage matter is found at the linearized level, since $\d T_{\mu\nu}[C](k) \to - T_{\mu\nu}(k)$ in the IR limit $k\to 0$. \harold{This suggests that linearized gravity effectively has reduced range. On the other hand, $\d T_{\mu\nu}[C]$ admits oscillatory Helmholz-type solutions $T_{00}[C] \sim m^2\frac{\cos(mr)}{r}$ around a point mass \cite{Steinacker:2026qzk}, and even solitonic solutions $T_{00}[C] \sim \frac{\sin(mr)}{r}$ without source. Hence the screening is mild and certainly not exponential as in massive gravity. Note that these effects arise from the interplay between Einstein-Hilbert and Yang-Mills term.}

 \harold{In the extreme IR regime, 
 the Yang-Mills action clearly dominates the E-H action, which has 2 more derivatives (cf. \eqref{variation-YM-C} together with \eqref{mirage-E-M-metric}. 
 Dropping the E-H term, the geometry is governed by a ''pre-gravity theory'' defined by the classical Yang-Mills term plus vacuum energy, where matter has only short-range effects because it is not charged.} That leads to the dynamical cosmological quantum spacetime $\cM^{3,1}$ discussed in section \ref{sec:dynamical-covar}, which is not governed by the Friedmann equations, and does not seem to require any fine-tuning. There is no need to invoke inflation to explain the homogeneity of the observed universe, and trans-Planckian issues \cite{Brandenberger:2012aj} are naturally resolved by the matrix framework.

It is tempting to conjecture that these IR modifications of gravity are (at least partially) responsible for so-called dark energy and possibly dark matter, marking the cross-over between the GR and YM regime. Some preliminary discussion of the IR modifications at the linearized level is given in \cite{Steinacker:2026qzk}, where halo-like effects are indeed observed;
however quantitative claims would be premature at this point\footnote{The validity of the linearized approximation is very limited, and a different behavior may arise at the non-linear level. Mirage matter also carries significant $\hs$ components, the meaning of which is obscure.}. In any case, the gravitational sector does contains an extra ''mirage'' sector
which goes beyond GR, displaying seemingly non-local dynamics if viewed in the language of GR.

\harold{Interesting tests  for this modified gravity regime should be provided by weak lensing observations such as \cite{Wenzl:2024sug}. Combined with the above modified Einstein equations, this would provide constraints on the mirage matter. The problem is of course that $L_{\rm cross}$ cannot be predicted at this point, and mirage matter is not yet understood in any detail. Nevertheless,
the above scale estimate in the weak-coupling regime $g_{\rm YM} < 1$ does imply new physics between $L_{\rm cosm}$ and $L_{\rm Pl}$.}

The combination of YM and EH terms also leads to extra dof, due to the 4-derivative structure of $S_{\rm EH}$ in terms of $\cA$ \cite{Steinacker:2026qzk}. 
The full dynamics then comprises not only the standard massless modes with dispersion $\Box_G = 0$ corresponding to gravitational waves as in GR, but also extra modes propagating at a reduced speed $\frac c{\sqrt{3}}$, with apparent tachyonic mass of order $-m^2$.
 These modes can also be seen in \eqref{lin-mod-Einstein-4} ff., where vacuum energy contributions are dropped.
They seem to be unstable at face value, however this is expected to be an artifact, and should be stabilized by taking into account vacuum energy \cite{Steinacker:2024unq}.
From the cosmological point of view, $S_{\rm EH}$ is a higher-derivative correction suppressed by the ''relatively high'' mass scale $m^2$ \eqref{m-cross-def-1} (from a cosmic point of view), as familiar from QFT. This  suggests that these extra modes are only relevant for the local physics (leading to gravity), but do not destroy the stability of the cosmological background.
Indeed  the cosmic background is found to be stabilized by vacuum energy \cite{Manta-new}, while the induced E-H term becomes irrelevant.
It may also be useful here to recall that the matrix model admits a conserved matrix energy momentum tensor $[\TT_a,\cT^{ab}] = 0$, which should keep potential instabilities in bound.

\section{Conclusion and outlook}

We have described how near-realistic quantum spacetime and gravity arise from the IKKT matrix models within a weak-coupling approach. The basic hypotheses of the underlying scenario  can now be justified to some degree,
including candidate backgrounds describing a 3+1 dimensional FLRW spacetime, stabilization of fuzzy extra dimensions, and the
emergence of an Einstein-Hilbert term. The semi-classical effective frame and metric with Minkowski signature are well understood in the weak coupling regime, leading to an understanding of gravity as a quantum effect on quantum spacetime.

In particular, covariant quantum spacetime provides a serious candidate for spacetime, leading to a consistent and appealing physical picture. While we focus on the $k=-1$ case, similar backgrounds with $k=0$ \cite{Gass:2025bqr} and $k=+1$ \cite{Steinacker:2017vqw} are also possible.
Some  understanding of why $\hs$ modes acquire a mass is offered in section \ref{sec:induced-masses}, so that (near-) realistic physics seems feasible on such a background.

There are many questions and tasks which remain to be addressed.
Clearly the most important problem is a detailed understanding of the gravitational dynamics. We have seen that this is highly nontrivial, since metric and frame are not fundamental but derived objects, which are moreover $\hs$-valued. The concept of local normal coordinates discussed in section \ref{sec:def-BG-LNC} is crucial to locally eliminate these $\hs$ components and make the connection to physics, which needs to be implemented and understood more generally. 
In particular, the ''mirage matter'' arising from the Yang-Mills flux and its physical significance needs to be better understood at the non-linear level. This is tied to the crossover scale $m$ \eqref{m-cross-def-1}, whose actual size depends on the background but needs better understanding.
Finally, stability of the cosmic background requires more detailed work taking into account vacuum energy  \cite{Manta-new}. This in turn is essential to understand the dynamics of the cosmic evolution and cosmic fluctuations. These and many other questions clearly require much more work, which is hopefully sparked and facilitated by the present review.

\subsection*{Acknowledgments}

This paper was sparked by lively discussions at the
''Large N Matrix Models and Emergent Geometry'' workshop at BIRS Banff, which is greatly appreciated. I would also like to thank the  collaborators of the recent papers underlying this overview, including Y. Asano, E. Battista, C. Gass, H. Kawai, K. Kumar, P-M Ho, A. Manta, T. Tran, as well as R. Brandenberger, C-S Chu, J. Karczmarek, among many others. 
This work is supported by the Austrian
Science Fund (FWF) grant P36479.

\bibliographystyle{JHEP}
\bibliography{twistor}

@article{Buckley-Bonanno:2026ygr,
    author = "Buckley-Bonanno, Samuel and Eckstein, Noah and Yelin, Susanne F.",
    title = "{Quantum simulation of gauge theories on dynamical spacetimes via Floquet-induced matrix models}",
    eprint = "2607.04040",
    archivePrefix = "arXiv",
    primaryClass = "quant-ph",
    month = "7",
    year = "2026"
}

@book{forrester2010log,
  title={Log-gases and random matrices (LMS-34)},
  author={Forrester, Peter J},
  year={2010},
  publisher={Princeton university press}
}

@article{Hrmo:2026ums,
    author = "Hrmo, M. and Kov{\'a}{\v{c}}ik, S. and Magdolenov{\'a}, K. and Manta, A. and Nede{\v{l}}kov{\'a}, K. and Rusn{\'a}k, P. and Steinacker, H. C. and Tekel, J.",
    title = "{Accretion, mergers, and metastability of fuzzy spheres in a three-matrix model}",
    eprint = "2609.00328",
    archivePrefix = "arXiv",
    primaryClass = "hep-th",
    month = "8",
    year = "2026"
}

@article{Liao:2025yfb,
    author = "Liao, Henry and Maeta, Reishi",
    title = "{New type of saddle in the Euclidean IKKT matrix model and its emergent geometry}",
    eprint = "2512.03161",
    archivePrefix = "arXiv",
    primaryClass = "hep-th",
    doi = "10.1103/gfcj-9vb3",
    journal = "Phys. Rev. D",
    volume = "114",
    number = "2",
    pages = "026021",
    year = "2026"
}

@article{Steinacker:2026jzp,
    author = "Steinacker, Harold C.",
    title = "{Quantum spacetime and quantum fluctuations in the IKKT model at weak coupling}",
    eprint = "2605.13294",
    archivePrefix = "arXiv",
    primaryClass = "hep-th",
    month = "5",
    year = "2026"
}

@article{Rinaldi:2021jbg,
    author = "Rinaldi, Enrico and Han, Xizhi and Hassan, Mohammad and Feng, Yuan and Nori, Franco and McGuigan, Michael and Hanada, Masanori",
    title = "{Matrix-Model Simulations Using Quantum Computing, Deep Learning, and Lattice Monte Carlo}",
    eprint = "2108.02942",
    archivePrefix = "arXiv",
    primaryClass = "quant-ph",
    reportNumber = "RIKEN-iTHEMS-Report-21, DMUS-MP-21/10",
    doi = "10.1103/PRXQuantum.3.010324",
    journal = "PRX Quantum",
    volume = "3",
    number = "1",
    pages = "010324",
    year = "2022"
}

@article{Sperling:2018hys,
    author = "Sperling, Marcus and Steinacker, Harold C.",
    title = "{Intersecting branes, Higgs sector, and chirality from $ \mathcal{N} $ = 4 SYM with soft SUSY breaking}",
    eprint = "1803.07323",
    archivePrefix = "arXiv",
    primaryClass = "hep-th",
    reportNumber = "UWThPh-2018-13, UWTHPH-2018-13",
    doi = "10.1007/JHEP04(2018)116",
    journal = "JHEP",
    volume = "04",
    pages = "116",
    year = "2018"
}

@article{Asano:2024def,
    author = "Asano, Yuhma and Nishimura, Jun and Piensuk, Worapat and Yamamori, Naoyuki",
    title = "{Defining the Type IIB Matrix Model without Breaking Lorentz Symmetry}",
    eprint = "2404.14045",
    archivePrefix = "arXiv",
    primaryClass = "hep-th",
    reportNumber = "UTHEP-787, KEK-TH-2617",
    doi = "10.1103/PhysRevLett.134.041603",
    journal = "Phys. Rev. Lett.",
    volume = "134",
    number = "4",
    pages = "041603",
    year = "2025"
}

@article{Alekseev:2000fd,
    author = "Alekseev, Anton Yu. and Recknagel, Andreas and Schomerus, Volker",
    title = "{Brane dynamics in background fluxes and noncommutative geometry}",
    eprint = "hep-th/0003187",
    archivePrefix = "arXiv",
    reportNumber = "IASSNS-HEP-00-23, AEI-2000-4",
    doi = "10.1088/1126-6708/2000/05/010",
    journal = "JHEP",
    volume = "05",
    pages = "010",
    year = "2000"
}

@article{Maeta:2026miu,
    author = "Maeta, Reishi",
    title = "{Regularized Master-Field Approximation for Large-$N$ Reduced Matrix Models}",
    eprint = "2605.10720",
    archivePrefix = "arXiv",
    primaryClass = "hep-th",
    month = "5",
    year = "2026"
}

@article{Maeta:2026oku,
    author = "Maeta, Reishi",
    title = "{Matrix bootstrap approximation without positivity constraint}",
    eprint = "2601.16099",
    archivePrefix = "arXiv",
    primaryClass = "hep-th",
    doi = "10.1007/JHEP05(2026)283",
    journal = "JHEP",
    volume = "05",
    pages = "283",
    year = "2026"
}

@article{Steinacker:2024huv,
    author = "Steinacker, Harold C. and Tran, Tung",
    title = "{Quantum $\mathfrak{hs}$-Yang-Mills from the IKKT matrix model}",
    eprint = "2405.09804",
    archivePrefix = "arXiv",
    primaryClass = "hep-th",
    month = "5",
    year = "2024"
}

@article{deWit:1988wri,
    author = "de Wit, B. and Hoppe, J. and Nicolai, H.",
    title = "{On the Quantum Mechanics of Supermembranes}",
    reportNumber = "THU-88-15, KA-THEP-6/88",
    doi = "10.1016/0550-3213(88)90116-2",
    journal = "Nucl. Phys. B",
    volume = "305",
    pages = "545",
    year = "1988"
}

@article{Ho:2025htr,
    author = "Ho, Pei-Ming and Kawai, Hikaru and Steinacker, Harold C.",
    title = "{General Relativity in IIB matrix model}",
    eprint = "2509.06646",
    archivePrefix = "arXiv",
    primaryClass = "hep-th",
    reportNumber = "UWThPh 2025-17, NITEP 258",
    doi = "10.1007/JHEP02(2026)070",
    journal = "JHEP",
    volume = "02",
    pages = "070",
    year = "2026"
}

@article{Maldacena:1997re,
    author = "Maldacena, Juan Martin",
    title = "{The Large $N$ limit of superconformal field theories and supergravity}",
    eprint = "hep-th/9711200",
    archivePrefix = "arXiv",
    reportNumber = "HUTP-97-A097, HUTP-98-A097",
    doi = "10.4310/ATMP.1998.v2.n2.a1",
    journal = "Adv. Theor. Math. Phys.",
    volume = "2",
    pages = "231--252",
    year = "1998"
}

@article{Brahma:2021tkh,
    author = "Brahma, Suddhasattwa and Brandenberger, Robert and Laliberte, Samuel",
    title = "{Emergent cosmology from matrix theory}",
    eprint = "2107.11512",
    archivePrefix = "arXiv",
    primaryClass = "hep-th",
    doi = "10.1007/JHEP03(2022)067",
    journal = "JHEP",
    volume = "03",
    pages = "067",
    year = "2022"
}

@article{Eguchi:1982nm,
    author = "Eguchi, Tohru and Kawai, Hikaru",
    title = "{Reduction of Dynamical Degrees of Freedom in the Large N Gauge Theory}",
    reportNumber = "UT-378-TOKYO",
    doi = "10.1103/PhysRevLett.48.1063",
    journal = "Phys. Rev. Lett.",
    volume = "48",
    pages = "1063",
    year = "1982"
}

@article{Doplicher:1994tu,
    author = "Doplicher, Sergio and Fredenhagen, Klaus and Roberts, John E.",
    title = "{The Quantum structure of space-time at the Planck scale and quantum fields}",
    eprint = "hep-th/0303037",
    archivePrefix = "arXiv",
    doi = "10.1007/BF02104515",
    journal = "Commun. Math. Phys.",
    volume = "172",
    pages = "187--220",
    year = "1995"
}

@article{Szabo:2001kg,
    author = "Szabo, Richard J.",
    title = "{Quantum field theory on noncommutative spaces}",
    eprint = "hep-th/0109162",
    archivePrefix = "arXiv",
    reportNumber = "HWM-01-35, EMPG-01-14",
    doi = "10.1016/S0370-1573(03)00059-0",
    journal = "Phys. Rept.",
    volume = "378",
    pages = "207--299",
    year = "2003"
}

@article{Eichhorn:2018yfc,
    author = "Eichhorn, Astrid",
    title = "{An asymptotically safe guide to quantum gravity and matter}",
    eprint = "1810.07615",
    archivePrefix = "arXiv",
    primaryClass = "hep-th",
    doi = "10.3389/fspas.2018.00047",
    journal = "Front. Astron. Space Sci.",
    volume = "5",
    pages = "47",
    year = "2019"
}

@article{Abanov:2025lln,
    author = "Abanov, Alexander G. and others",
    title = "{Quantum Geometry of Data}",
    eprint = "2507.21135",
    archivePrefix = "arXiv",
    primaryClass = "cs.LG",
    month = "7",
    year = "2025"
}

@book{Madore:2000aq,
    author = "Madore, J.",
    title = "{An introduction to noncommutative differential geometry and its physical applications}",
    publisher = "Cambridge University Press",
    volume = "257",
    year = "2000"
}

@book{Percacci:2017fkn,
    author = "Percacci, Robert",
    title = "{An Introduction to Covariant Quantum Gravity and Asymptotic Safety}",
    doi = "10.1142/10369",
    isbn = "978-981-320-717-2, 978-981-320-719-6",
    publisher = "World Scientific",
    series = "100 Years of General Relativity",
    volume = "3",
    year = "2017"
}

@article{Douglas:2001ba,
    author = "Douglas, Michael R. and Nekrasov, Nikita A.",
    title = "{Noncommutative field theory}",
    eprint = "hep-th/0106048",
    archivePrefix = "arXiv",
    reportNumber = "ITEP-TH-31-01, IHES-P-01-27, RUNHETC-2001-18",
    doi = "10.1103/RevModPhys.73.977",
    journal = "Rev. Mod. Phys.",
    volume = "73",
    pages = "977--1029",
    year = "2001"
}

@article{Grosse:1995ar,
    author = "Grosse, H. and Klimcik, C. and Presnajder, P.",
    title = "{Towards finite quantum field theory in noncommutative geometry}",
    eprint = "hep-th/9505175",
    archivePrefix = "arXiv",
    reportNumber = "CERN-TH-95-138, UWTHPH-19-1995",
    doi = "10.1007/BF02083810",
    journal = "Int. J. Theor. Phys.",
    volume = "35",
    pages = "231--244",
    year = "1996"
}

@article{Gonzalez-Arroyo:1982hyq,
    author = "Gonzalez-Arroyo, Antonio and Okawa, M.",
    title = "{The Twisted Eguchi-Kawai Model: A Reduced Model for Large N Lattice Gauge Theory}",
    reportNumber = "BNL-32393",
    doi = "10.1103/PhysRevD.27.2397",
    journal = "Phys. Rev. D",
    volume = "27",
    pages = "2397",
    year = "1983"
}

@article{Brahma:2022dsd,
    author = "Brahma, Suddhasattwa and Brandenberger, Robert and Laliberte, Samuel",
    title = "{Emergent metric space-time from matrix theory}",
    eprint = "2206.12468",
    archivePrefix = "arXiv",
    primaryClass = "hep-th",
    doi = "10.1007/JHEP09(2022)031",
    journal = "JHEP",
    volume = "09",
    pages = "031",
    year = "2022"
}

@article{Brandenberger:2024ddi,
    author = "Brandenberger, Robert and Pasiecznik, Julia",
    title = "{Origin of the SO(9){\textrightarrow}SO(3){\texttimes}SO(6) symmetry breaking in the type IIB matrix model}",
    eprint = "2409.00254",
    archivePrefix = "arXiv",
    primaryClass = "hep-th",
    doi = "10.1103/bqzp-n28g",
    journal = "Phys. Rev. D",
    volume = "112",
    number = "2",
    pages = "026006",
    year = "2025"
}

@article{Koch:2021yeb,
    author = "Koch, Robert de Mello and Jevicki, Antal and Liu, Xianlong and Mathaba, Kagiso and Rodrigues, Jo{\~a}o P.",
    title = "{Large N optimization for multi-matrix systems}",
    eprint = "2108.08803",
    archivePrefix = "arXiv",
    primaryClass = "hep-th",
    doi = "10.1007/JHEP01(2022)168",
    journal = "JHEP",
    volume = "01",
    pages = "168",
    year = "2022"
}

@article{Banks:1996vh,
    author = "Banks, Tom and Fischler, W. and Shenker, S. H. and Susskind, Leonard",
    title = "{M theory as a matrix model: A conjecture}",
    eprint = "hep-th/9610043",
    archivePrefix = "arXiv",
    reportNumber = "RU-96-95, SU-ITP-96-12, UTTG-13-96",
    doi = "10.1103/PhysRevD.55.5112",
    journal = "Phys. Rev. D",
    volume = "55",
    pages = "5112--5128",
    year = "1997"
}

@article{Loll:1998aj,
    author = "Loll, Renate",
    title = "{Discrete approaches to quantum gravity in four-dimensions}",
    eprint = "gr-qc/9805049",
    archivePrefix = "arXiv",
    reportNumber = "AEI-063A, AEI-063",
    doi = "10.12942/lrr-1998-13",
    journal = "Living Rev. Rel.",
    volume = "1",
    pages = "13",
    year = "1998"
}

@article{Loll:2019rdj,
    author = "Loll, R.",
    title = "{Quantum Gravity from Causal Dynamical Triangulations: A Review}",
    eprint = "1905.08669",
    archivePrefix = "arXiv",
    primaryClass = "hep-th",
    doi = "10.1088/1361-6382/ab57c7",
    journal = "Class. Quant. Grav.",
    volume = "37",
    number = "1",
    pages = "013002",
    year = "2020"
}

@article{Perez:2012wv,
    author = "Perez, Alejandro",
    title = "{The Spin Foam Approach to Quantum Gravity}",
    eprint = "1205.2019",
    archivePrefix = "arXiv",
    primaryClass = "gr-qc",
    doi = "10.12942/lrr-2013-3",
    journal = "Living Rev. Rel.",
    volume = "16",
    pages = "3",
    year = "2013"
}

@article{Rovelli:2008zza,
    author = "Rovelli, Carlo",
    title = "{Loop quantum gravity}",
    doi = "10.12942/lrr-2008-5",
    journal = "Living Rev. Rel.",
    volume = "11",
    pages = "5",
    year = "2008"
}

@article{Grana:2005jc,
    author = "Grana, Mariana",
    title = "{Flux compactifications in string theory: A Comprehensive review}",
    eprint = "hep-th/0509003",
    archivePrefix = "arXiv",
    reportNumber = "LPTENS-05-26, CPHT-RR-049-0805",
    doi = "10.1016/j.physrep.2005.10.008",
    journal = "Phys. Rept.",
    volume = "423",
    pages = "91--158",
    year = "2006"
}

@article{Susskind:2003kw,
    author = "Susskind, Leonard",
    editor = "Carr, Bernard J.",
    title = "{The Anthropic landscape of string theory}",
    eprint = "hep-th/0302219",
    archivePrefix = "arXiv",
    pages = "247--266",
    month = "2",
    year = "2003"
}

@article{Wenzl:2024sug,
    author = "Wenzl, Lukas and Bean, Rachel and Chen, Shi-Fan and Farren, Gerrit S. and Madhavacheril, Mathew S. and Marques, Gabriela A. and Qu, Frank J. and Sehgal, Neelima and Sherwin, Blake D. and van Engelen, Alexander",
    title = "{Constraining gravity with a new precision EG estimator using Planck + SDSS BOSS data}",
    eprint = "2401.12971",
    archivePrefix = "arXiv",
    primaryClass = "astro-ph.CO",
    reportNumber = "FERMILAB-PUB-24-0061-PPD",
    doi = "10.1103/PhysRevD.109.083540",
    journal = "Phys. Rev. D",
    volume = "109",
    number = "8",
    pages = "083540",
    year = "2024"
}

@article{Jack:2001cr,
    author = "Jack, I. and Jones, D. R. T.",
    title = "{Ultraviolet finiteness in noncommutative supersymmetric theories}",
    eprint = "hep-th/0109195",
    archivePrefix = "arXiv",
    reportNumber = "LTH-519",
    doi = "10.1088/1367-2630/3/1/319",
    journal = "New J. Phys.",
    volume = "3",
    pages = "19",
    year = "2001"
}

@article{Fukuma:1997en,
    author = "Fukuma, Masafumi and Kawai, Hikaru and Kitazawa, Yoshihisa and Tsuchiya, Asato",
    title = "{String field theory from IIB matrix model}",
    eprint = "hep-th/9705128",
    archivePrefix = "arXiv",
    reportNumber = "KEK-TH-518, TIT-HEP-369",
    doi = "10.1016/S0550-3213(97)00584-1",
    journal = "Nucl. Phys. B",
    volume = "510",
    pages = "158--174",
    year = "1998"
}

@article{Aoki:1998bq,
    author = "Aoki, Hajime and Iso, Satoshi and Kawai, Hikaru and Kitazawa, Yoshihisa and Tsuchiya, Asato and Tada, Tsukasa",
    editor = "Iso, S. and Kawai, H. and Natsuume, M.",
    title = "{IIB matrix model}",
    eprint = "hep-th/9908038",
    archivePrefix = "arXiv",
    reportNumber = "KEK-TH-635",
    doi = "10.1143/PTPS.134.47",
    journal = "Prog. Theor. Phys. Suppl.",
    volume = "134",
    pages = "47--83",
    year = "1999"
}

@article{Steinacker:2022kji,
    author = "Steinacker, Harold C. and Tekel, Juraj",
    title = "{String modes, propagators and loops on fuzzy spaces}",
    eprint = "2203.02376",
    archivePrefix = "arXiv",
    primaryClass = "hep-th",
    reportNumber = "UWThPh-2022-3",
    doi = "10.1007/JHEP06(2022)136",
    journal = "JHEP",
    volume = "06",
    pages = "136",
    year = "2022"
}

@book{Steinacker:2024unq,
    author = "Steinacker, Harold C.",
    title = "{Quantum Geometry, Matrix Theory, and Gravity}",
    doi = "10.1017/9781009440776",
    isbn = "978-1-00-944077-6, 978-1-00-944078-3",
    publisher = "Cambridge University Press",
    month = "4",
    year = "2024"
}

@article{Nishimura:2019qal,
    author = "Nishimura, Jun and Tsuchiya, Asato",
    title = "{Complex Langevin analysis of the space-time structure in the Lorentzian type IIB matrix model}",
    eprint = "1904.05919",
    archivePrefix = "arXiv",
    primaryClass = "hep-th",
    reportNumber = "KEK-TH-2119",
    doi = "10.1007/JHEP06(2019)077",
    journal = "JHEP",
    volume = "06",
    pages = "077",
    year = "2019"
}

@article{Battista:2023glw,
    author = "Battista, Emmanuele and Steinacker, Harold C.",
    title = "{One-loop effective action of the IKKT model for cosmological backgrounds}",
    eprint = "2310.11126",
    archivePrefix = "arXiv",
    primaryClass = "hep-th",
    month = "10",
    year = "2023"
}

@article{Kumar:2023bxg,
    author = "Kumar, Kaushlendra and Steinacker, Harold C.",
    title = "{Modified Einstein equations from the 1-loop effective action of the IKKT model}",
    eprint = "2312.01317",
    archivePrefix = "arXiv",
    primaryClass = "hep-th",
    month = "12",
    year = "2023"
}

@article{Kugo:1979gm,
    author = "Kugo, Taichiro and Ojima, Izumi",
    title = "{Local Covariant Operator Formalism of Nonabelian Gauge Theories and Quark Confinement Problem}",
    reportNumber = "KUNS-493",
    doi = "10.1143/PTPS.66.1",
    journal = "Prog. Theor. Phys. Suppl.",
    volume = "66",
    pages = "1--130",
    year = "1979"
}

@article{Gass:2025bqr,
    author = "Ga{\ss}, Christian and Steinacker, Harold C.",
    title = "{Spatially flat cosmological quantum spacetimes}",
    eprint = "2510.21283",
    archivePrefix = "arXiv",
    primaryClass = "hep-th",
    doi = "10.1103/6l9d-njcr",
    journal = "Phys. Rev. D",
    volume = "113",
    number = "2",
    pages = "024050",
    year = "2026"
}

@article{Aschieri:2006uw,
    author = "Aschieri, Paolo and Grammatikopoulos, Theodoros and Steinacker, Harold and Zoupanos, George",
    title = "{Dynamical generation of fuzzy extra dimensions, dimensional reduction and symmetry breaking}",
    eprint = "hep-th/0606021",
    archivePrefix = "arXiv",
    reportNumber = "UWTHPH-2006-12, DISTA-UPO-06",
    doi = "10.1088/1126-6708/2006/09/026",
    journal = "JHEP",
    volume = "09",
    pages = "026",
    year = "2006"
}

@article{Ishibashi:1996xs,
    author = "Ishibashi, N. and Kawai, H. and Kitazawa, Y. and Tsuchiya, A.",
    title = "{A Large N reduced model as superstring}",
    eprint = "hep-th/9612115",
    archivePrefix = "arXiv",
    reportNumber = "KEK-TH-503",
    doi = "10.1016/S0550-3213(97)00290-3",
    journal = "Nucl. Phys. B",
    volume = "498",
    pages = "467--491",
    year = "1997"
}

@article{Steinacker:2026qzk,
    author = "Steinacker, Harold C.",
    title = "{Modified gravity at large scales on quantum spacetime in the IKKT model}",
    eprint = "2601.08031",
    archivePrefix = "arXiv",
    primaryClass = "hep-th",
    doi = "10.1007/JHEP04(2026)044",
    journal = "JHEP",
    volume = "04",
    pages = "044",
    year = "2026"
}

@article{Steinacker:2017vqw,
    author = "Steinacker, Harold C.",
    title = "{Cosmological space-times with resolved Big Bang in Yang-Mills matrix models}",
    eprint = "1709.10480",
    archivePrefix = "arXiv",
    primaryClass = "hep-th",
    reportNumber = "UWTHPH-2017-31",
    doi = "10.1007/JHEP02(2018)033",
    journal = "JHEP",
    volume = "02",
    pages = "033",
    year = "2018"
}

@article{Castelino:1997rv,
    author = "Castelino, Judith and Lee, Sangmin and Taylor, Washington",
    title = "{Longitudinal five-branes as four spheres in matrix theory}",
    eprint = "hep-th/9712105",
    archivePrefix = "arXiv",
    reportNumber = "PUPT-1749",
    doi = "10.1016/S0550-3213(98)00291-0",
    journal = "Nucl. Phys. B",
    volume = "526",
    pages = "334--350",
    year = "1998"
}

@article{Steinacker:2020nva,
    author = "Steinacker, Harold C.",
    title = "{Quantum (Matrix) Geometry and Quasi-Coherent States}",
    eprint = "2009.03400",
    archivePrefix = "arXiv",
    primaryClass = "hep-th",
    reportNumber = "UWThPh-2020-22",
    doi = "10.1088/1751-8121/abd735",
    journal = "J. Phys. A",
    volume = "54",
    number = "5",
    pages = "055401",
    year = "2021"
}

@article{Blaschke:2011qu,
    author = "Blaschke, Daniel N. and Steinacker, Harold",
    title = "{On the 1-loop effective action for the IKKT model and non-commutative branes}",
    eprint = "1109.3097",
    archivePrefix = "arXiv",
    primaryClass = "hep-th",
    reportNumber = "UWTHPH-2011-29",
    doi = "10.1007/JHEP10(2011)120",
    journal = "JHEP",
    volume = "10",
    pages = "120",
    year = "2011"
}

@article{Sperling:2018xrm,
    author = "Sperling, Marcus and Steinacker, Harold C.",
    title = "{The fuzzy 4-hyperboloid $H^4_n$ and higher-spin in Yang\textendash{}Mills matrix models}",
    eprint = "1806.05907",
    archivePrefix = "arXiv",
    primaryClass = "hep-th",
    reportNumber = "UWThPh-2018-16, UWTHPH-2018-16",
    doi = "10.1016/j.nuclphysb.2019.02.027",
    journal = "Nucl. Phys. B",
    volume = "941",
    pages = "680--743",
    year = "2019"
}

@article{Sperling:2019xar,
    author = "Sperling, Marcus and Steinacker, Harold C.",
    title = "{Covariant cosmological quantum space-time, higher-spin and gravity in the IKKT matrix model}",
    eprint = "1901.03522",
    archivePrefix = "arXiv",
    primaryClass = "hep-th",
    reportNumber = "UWThPh-2019-02",
    doi = "10.1007/JHEP07(2019)010",
    journal = "JHEP",
    volume = "07",
    pages = "010",
    year = "2019"
}

@article{Steinacker:2019awe,
    author = "Steinacker, Harold C.",
    title = "{Higher-spin kinematics \& no ghosts on quantum space-time in Yang-Mills matrix models}",
    eprint = "1910.00839",
    archivePrefix = "arXiv",
    primaryClass = "hep-th",
    reportNumber = "UWThPh-2019-28",
    month = "10",
    year = "2019"
}

@article{Fernando:2009fq,
    author = "Fernando, Sudarshan and Gunaydin, Murat",
    title = "{Minimal unitary representation of SU(2,2) and its deformations as massless conformal fields and their supersymmetric extensions}",
    eprint = "0908.3624",
    archivePrefix = "arXiv",
    primaryClass = "hep-th",
    doi = "10.1063/1.3447773",
    journal = "J. Math. Phys.",
    volume = "51",
    pages = "082301",
    year = "2010"
}

@article{Austing:2001bd,
    author = "Austing, Peter and Wheater, John F.",
    title = "{The Convergence of Yang-Mills integrals}",
    eprint = "hep-th/0101071",
    archivePrefix = "arXiv",
    reportNumber = "OUTP-01-01P",
    doi = "10.1088/1126-6708/2001/02/028",
    journal = "JHEP",
    volume = "02",
    pages = "028",
    year = "2001"
}

@article{Krauth:1998yu,
    author = "Krauth, Werner and Staudacher, Matthias",
    title = "{Finite Yang-Mills integrals}",
    eprint = "hep-th/9804199",
    archivePrefix = "arXiv",
    reportNumber = "AEI-063",
    doi = "10.1016/S0370-2693(98)00814-4",
    journal = "Phys. Lett. B",
    volume = "435",
    pages = "350--355",
    year = "1998"
}

@article{Steinacker:2019fcb,
    author = "Steinacker, Harold C.",
    title = "{On the quantum structure of space-time, gravity, and higher spin in matrix models}",
    eprint = "1911.03162",
    archivePrefix = "arXiv",
    primaryClass = "hep-th",
    reportNumber = "UWThPh-2019-32",
    doi = "10.1088/1361-6382/ab857f",
    journal = "Class. Quant. Grav.",
    volume = "37",
    number = "11",
    pages = "113001",
    year = "2020"
}

@article{Steinacker:2020xph,
    author = "Steinacker, Harold C.",
    title = "{Higher-spin gravity and torsion on quantized space-time in matrix models}",
    eprint = "2002.02742",
    archivePrefix = "arXiv",
    primaryClass = "hep-th",
    reportNumber = "UWThPh-2020-5",
    doi = "10.1007/JHEP04(2020)111",
    journal = "JHEP",
    volume = "04",
    pages = "111",
    year = "2020"
}

@article{Fredenhagen:2021bnw,
    author = "Fredenhagen, Stefan and Steinacker, Harold C.",
    title = "{Exploring the gravity sector of emergent higher-spin gravity: effective action and a solution}",
    eprint = "2101.07297",
    archivePrefix = "arXiv",
    primaryClass = "hep-th",
    reportNumber = "UWThPh 2021-1",
    doi = "10.1007/JHEP05(2021)183",
    journal = "JHEP",
    volume = "05",
    pages = "183",
    year = "2021"
}

@article{Kawai:2002jk,
    author = "Kawai, H. and Kawamoto, Shoichi and Kuroki, Tsunehide and Matsuo, T. and Shinohara, S.",
    title = "{Mean field approximation of IIB matrix model and emergence of four-dimensional space-time}",
    eprint = "hep-th/0204240",
    archivePrefix = "arXiv",
    reportNumber = "KUNS-1779",
    doi = "10.1016/S0550-3213(02)00908-2",
    journal = "Nucl. Phys. B",
    volume = "647",
    pages = "153--189",
    year = "2002"
}

@article{Steinacker:2010rh,
    author = "Steinacker, Harold",
    title = "{Emergent Geometry and Gravity from Matrix Models: an Introduction}",
    eprint = "1003.4134",
    archivePrefix = "arXiv",
    primaryClass = "hep-th",
    reportNumber = "UWTHPH-2010-4",
    doi = "10.1088/0264-9381/27/13/133001",
    journal = "Class. Quant. Grav.",
    volume = "27",
    pages = "133001",
    year = "2010"
}

@book{connes1994noncommutative,
  title={Noncommutative Geometry},
  author={Connes, Alain},
  year={1994},
  publisher={Academic Press},
  address={San Diego, CA},
  isbn={978-0121858605}
}

@article{Steinacker:2021yxt,
    author = "Steinacker, Harold C.",
    title = "{Gravity as a quantum effect on quantum space-time}",
    eprint = "2110.03936",
    archivePrefix = "arXiv",
    primaryClass = "hep-th",
    reportNumber = "UWTHPh-2021-17",
    doi = "10.1016/j.physletb.2022.136946",
    journal = "Phys. Lett. B",
    volume = "827",
    pages = "136946",
    year = "2022"
}

@article{Battista:2022hqn,
    author = "Battista, Emmanuele and Steinacker, Harold C.",
    title = "{On the propagation across the big bounce in an open quantum FLRW cosmology}",
    eprint = "2207.01295",
    archivePrefix = "arXiv",
    primaryClass = "gr-qc",
    reportNumber = "UWThPh-2022-10",
    doi = "10.1140/epjc/s10052-022-10874-0",
    journal = "Eur. Phys. J. C",
    volume = "82",
    number = "10",
    pages = "909",
    year = "2022"
}

@article{Karczmarek:2022ejn,
    author = "Karczmarek, Joanna L. and Steinacker, Harold C.",
    title = "{Cosmic time evolution and propagator from a Yang-Mills matrix model}",
    eprint = "2207.00399",
    archivePrefix = "arXiv",
    primaryClass = "hep-th",
    reportNumber = "UWThPh-2022-8",
    month = "7",
    year = "2022"
}

@article{Steinacker:2023myp,
    author = "Steinacker, Harold C.",
    title = "{One-loop effective action and emergent gravity on quantum spaces in the IKKT matrix model}",
    eprint = "2303.08012",
    archivePrefix = "arXiv",
    primaryClass = "hep-th",
    reportNumber = "UWThPh-2023-9",
    month = "3",
    year = "2023"
}

@article{Iso:2001mg,
    author = "Iso, Satoshi and Kimura, Yusuke and Tanaka, Kanji and Wakatsuki, Kazunori",
    title = "{Noncommutative gauge theory on fuzzy sphere from matrix model}",
    eprint = "hep-th/0101102",
    archivePrefix = "arXiv",
    reportNumber = "KEK-TH-737, TIT-HEP-462",
    doi = "10.1016/S0550-3213(01)00173-0",
    journal = "Nucl. Phys. B",
    volume = "604",
    pages = "121--147",
    year = "2001"
}

@article{Delgadillo-Blando:2007mqd,
    author = "Delgadillo-Blando, Rodrigo and O'Connor, Denjoe and Ydri, Badis",
    title = "{Geometry in Transition: A Model of Emergent Geometry}",
    eprint = "0712.3011",
    archivePrefix = "arXiv",
    primaryClass = "hep-th",
    reportNumber = "DIAS-PREPRINT-07-22",
    doi = "10.1103/PhysRevLett.100.201601",
    journal = "Phys. Rev. Lett.",
    volume = "100",
    pages = "201601",
    year = "2008"
}

@article{Delgadillo-Blando:2008cuz,
    author = "Delgadillo-Blando, Rodrigo and O'Connor, Denjoe and Ydri, Badis",
    title = "{Matrix Models, Gauge Theory and Emergent Geometry}",
    eprint = "0806.0558",
    archivePrefix = "arXiv",
    primaryClass = "hep-th",
    doi = "10.1088/1126-6708/2009/05/049",
    journal = "JHEP",
    volume = "05",
    pages = "049",
    year = "2009"
}

@article{Azuma:2005bj,
    author = "Azuma, Takehiro and Bal, Subrata and Nishimura, Jun",
    title = "{Dynamical generation of gauge groups in the massive Yang-Mills-Chern-Simons matrix model}",
    eprint = "hep-th/0504217",
    archivePrefix = "arXiv",
    reportNumber = "KEK-TH-1014, RIKEN-TH-41",
    doi = "10.1103/PhysRevD.72.066005",
    journal = "Phys. Rev. D",
    volume = "72",
    pages = "066005",
    year = "2005"
}

@article{Seiberg:1999vs,
    author = "Seiberg, Nathan and Witten, Edward",
    title = "{String theory and noncommutative geometry}",
    eprint = "hep-th/9908142",
    archivePrefix = "arXiv",
    reportNumber = "IASSNS-HEP-99-74",
    doi = "10.1088/1126-6708/1999/09/032",
    journal = "JHEP",
    volume = "09",
    pages = "032",
    year = "1999"
}

@article{Connes:1997cr,
    author = "Connes, Alain and Douglas, Michael R. and Schwarz, Albert S.",
    title = "{Noncommutative geometry and matrix theory: Compactification on tori}",
    eprint = "hep-th/9711162",
    archivePrefix = "arXiv",
    reportNumber = "RU-97-94",
    doi = "10.1088/1126-6708/1998/02/003",
    journal = "JHEP",
    volume = "02",
    pages = "003",
    year = "1998"
}

@article{Anagnostopoulos:2026qvz,
    author = "Anagnostopoulos, Konstantinos N. and Azuma, Takehiro and Hirasawa, Mitsuaki and Nishimura, Jun and Papadoudis, Stratos and Tsuchiya, Asato",
    title = "{The emergence of (3+1)-dimensional expanding spacetime from complex Langevin simulations of the Lorentzian type IIB matrix model with deformations}",
    eprint = "2604.19836",
    archivePrefix = "arXiv",
    primaryClass = "hep-th",
    reportNumber = "KEK-TH-2826",
    month = "4",
    year = "2026"
}

@article{Anagnostopoulos:2022dak,
    author = "Anagnostopoulos, Konstantinos N. and Azuma, Takehiro and Hatakeyama, Kohta and Hirasawa, Mitsuaki and Ito, Yuta and Nishimura, Jun and Papadoudis, Stratos Kovalkov and Tsuchiya, Asato",
    title = "{Progress in the numerical studies of the type IIB matrix model}",
    eprint = "2210.17537",
    archivePrefix = "arXiv",
    primaryClass = "hep-th",
    reportNumber = "KEK-TH-2470",
    doi = "10.1140/epjs/s11734-023-00849-x",
    journal = "Eur. Phys. J. ST",
    volume = "232",
    number = "23-24",
    pages = "3681--3695",
    year = "2023"
}

@article{Anagnostopoulos:2026utg,
    author = "Anagnostopoulos, Konstantinos N. and Azuma, Takehiro and Hirasawa, Mitsuaki and Nishimura, Jun and Tsuchiya, Asato and Yamamori, Naoyuki",
    title = "{Impact of supersymmetry on the dynamical emergence of the spacetime in the type IIB matrix model with the Lorentz symmetry ''gauge fixed''}",
    eprint = "2604.25564",
    archivePrefix = "arXiv",
    primaryClass = "hep-lat",
    reportNumber = "KEK-TH 2806",
    month = "4",
    year = "2026"
}

@article{Hatakeyama:2019jyw,
    author = "Hatakeyama, Kohta and Matsumoto, Akira and Nishimura, Jun and Tsuchiya, Asato and Yosprakob, Atis",
    title = "{The emergence of expanding space{\textendash}time and intersecting D-branes from classical solutions in the Lorentzian type IIB matrix model}",
    eprint = "1911.08132",
    archivePrefix = "arXiv",
    primaryClass = "hep-th",
    reportNumber = "KEK-TH-2169",
    doi = "10.1093/ptep/ptaa042",
    journal = "PTEP",
    volume = "2020",
    number = "4",
    pages = "043B10",
    year = "2020"
}

@article{Chepelev:1997av,
    author = "Chepelev, I. and Tseytlin, Arkady A.",
    title = "{Interactions of type IIB D-branes from D instanton matrix model}",
    eprint = "hep-th/9705120",
    archivePrefix = "arXiv",
    reportNumber = "ITEP-TH-19-97, IMPERIAL-TP-96-97-47",
    doi = "10.1016/S0550-3213(97)00658-5",
    journal = "Nucl. Phys. B",
    volume = "511",
    pages = "629--646",
    year = "1998"
}

@article{Bonelli:2002mb,
    author = "Bonelli, Giulio",
    title = "{Matrix strings in pp wave backgrounds from deformed superYang-Mills theory}",
    eprint = "hep-th/0205213",
    archivePrefix = "arXiv",
    reportNumber = "ULB-TH-02-15",
    doi = "10.1088/1126-6708/2002/08/022",
    journal = "JHEP",
    volume = "08",
    pages = "022",
    year = "2002"
}

@article{Ciceri:2025wpb,
    author = "Ciceri, Franz and Samtleben, Henning",
    title = "{Supergravity dual for Ishibashi-Kawai-Kitazawa-Tsuchiya holography}",
    eprint = "2511.23111",
    archivePrefix = "arXiv",
    primaryClass = "hep-th",
    doi = "10.1103/gmhl-mmg5",
    journal = "Phys. Rev. D",
    volume = "113",
    number = "4",
    pages = "046001",
    year = "2026"
}

@article{Aoyama:2010ry,
    author = "Aoyama, Tatsumi and Nishimura, Jun and Okubo, Toshiyuki",
    title = "{Spontaneous breaking of the rotational symmetry in dimensionally reduced super Yang-Mills models}",
    eprint = "1007.0883",
    archivePrefix = "arXiv",
    primaryClass = "hep-th",
    doi = "10.1143/PTP.125.537",
    journal = "Prog. Theor. Phys.",
    volume = "125",
    pages = "537--563",
    year = "2011"
}

@article{Manta-new,
    author = "Manta, Alessandro and Steinacker, Harold C.",
    title = "{Dynamical Covariant Quantum Spacetimes in IKKT stabilized at
one-loop}",
 journal = "in preparation",
 year = "2026"
}

@article{Brandenberger:2012aj,
    author = "Brandenberger, Robert H. and Martin, Jerome",
    title = "{Trans-Planckian Issues for Inflationary Cosmology}",
    eprint = "1211.6753",
    archivePrefix = "arXiv",
    primaryClass = "astro-ph.CO",
    doi = "10.1088/0264-9381/30/11/113001",
    journal = "Class. Quant. Grav.",
    volume = "30",
    pages = "113001",
    year = "2013"
}

@article{Manta:2025inq,
    author = "Manta, Alessandro and Steinacker, Harold C.",
    title = "{Minimal covariant quantum space-time}",
    eprint = "2502.02498",
    archivePrefix = "arXiv",
    primaryClass = "hep-th",
    reportNumber = "UWThPh 2025-4",
    doi = "10.1088/1751-8121/adcc6e",
    journal = "J. Phys. A",
    volume = "58",
    number = "17",
    pages = "175204",
    year = "2025"
}

@article{Manta:2024vol,
    author = "Manta, Alessandro and Steinacker, Harold C. and Tran, Tung",
    title = "{$ \mathfrak{hs} $-extended gravity from the IKKT matrix model}",
    eprint = "2411.02598",
    archivePrefix = "arXiv",
    primaryClass = "hep-th",
    doi = "10.1007/JHEP02(2025)031",
    journal = "JHEP",
    volume = "02",
    pages = "031",
    year = "2025"
}

@article{Steinacker:2014eua,
    author = "Steinacker, Harold C.",
    title = "{Spinning squashed extra dimensions and chiral gauge theory from N=4 SYM}",
    eprint = "1411.3139",
    archivePrefix = "arXiv",
    primaryClass = "hep-th",
    reportNumber = "UWTHPH-2014-25",
    doi = "10.1016/j.nuclphysb.2015.04.023",
    journal = "Nucl. Phys. B",
    volume = "896",
    pages = "212--243",
    year = "2015"
}

@article{Iso:2015mva,
    author = "Iso, Satoshi and Kitazawa, Noriaki",
    title = "{Revolving D-branes and Spontaneous Gauge Symmetry Breaking}",
    eprint = "1507.04834",
    archivePrefix = "arXiv",
    primaryClass = "hep-ph",
    reportNumber = "KEK-TH-1849",
    doi = "10.1093/ptep/ptv157",
    journal = "PTEP",
    volume = "2015",
    number = "12",
    pages = "123B01",
    year = "2015"
}

@article{Hartnoll:2024csr,
    author = "Hartnoll, Sean A. and Liu, Jun",
    title = "{The polarised IKKT matrix model}",
    eprint = "2409.18706",
    archivePrefix = "arXiv",
    primaryClass = "hep-th",
    doi = "10.1007/JHEP03(2025)060",
    journal = "JHEP",
    volume = "03",
    pages = "060",
    year = "2025"
}

@article{Komatsu:2024bop,
    author = "Komatsu, Shota and Martina, Adrien and Penedones, Jo{\~a}o and Vuignier, Antoine and Zhao, Xiang",
    title = "{Einstein gravity from a matrix integral -- Part I}",
    eprint = "2410.18173",
    archivePrefix = "arXiv",
    primaryClass = "hep-th",
    month = "10",
    year = "2024"
}

@article{Komatsu:2024ydh,
    author = "Komatsu, Shota and Martina, Adrien and Penedones, Joao and Vuignier, Antoine and Zhao, Xiang",
    title = "{Einstein gravity from a matrix integral -- Part II}",
    eprint = "2411.18678",
    archivePrefix = "arXiv",
    primaryClass = "hep-th",
    month = "11",
    year = "2024"
}

@article{Manta:2025tcl,
    author = "Manta, Alessandro and Steinacker, Harold C.",
    title = "{Dynamical Covariant Quantum Spacetime with Fuzzy Extra Dimensions in the IKKT model}",
    eprint = "2509.24753",
    archivePrefix = "arXiv",
    primaryClass = "hep-th",
    month = "9",
    year = "2025"
}

@article{Matusis:2000jf,
  author = {Matusis, Alec and Susskind, Leonard and Toumbas, Nicolaos},
  title = {{The IR / UV connection in the noncommutative gauge theories}},
  journal = {JHEP},
  volume = {12},
  year = {2000},
  pages = {002},
  doi = {10.1088/1126-6708/2000/12/002},
  eprint = {hep-th/0002075},
  archiveprefix = {arXiv},
  primaryclass = {hep-th},
  reportnumber = {SU-ITP-00-07},
  slaccitation = {%%CITATION = HEP-TH/0002075;%%},
}

\end{document}